\documentclass[sigconf]{acmart}

\AtBeginDocument{%
  \providecommand\BibTeX{{%
    \normalfont B\kern-0.5em{\scshape i\kern-0.25em b}\kern-0.8em\TeX}}}

\setcopyright{acmcopyright}
\copyrightyear{2026}
\acmYear{2026}
\acmDOI{10.1145/3772318.3791283}

\acmBooktitle{Proceedings of the 2026 CHI Conference on Human Factors in Computing Systems (CHI ’26)} 
\acmPrice{15.00}
\acmISBN{979-8-4007-2278-3/26/04}

\usepackage{enumitem}
\usepackage{xspace}

\usepackage[capitalise, noabbrev]{cleveref}
\usepackage{subcaption}
\usepackage{multirow}
\usepackage{makecell}
\usepackage{colortbl}
\usepackage{listings}
\usepackage{framed}

\newcommand{\paragraphBold}[1]{\paragraph{\emph{\textbf{#1}}}\xspace}

\newcommand{\code}[1]{\texttt{#1}\xspace}
\newcommand{\step}[1]{\texttt{\emph{#1}}\xspace}

\definecolor{cquote}{HTML}{3c4043}
\newcommand{\quoteinline}[1]{{\color{cquote}\emph{``#1'’}\xspace}}
\newcommand{\citequote}[2]{``#2'' (#1)}

\begin{document}

\title[Teaching Effective AI Use in Computational Data Analysis]{Not Everyone Wins with LLMs: Behavioral Patterns and Pedagogical Implications for AI Literacy in Programmatic Data Science}

\author{Qianou Ma}
\email{qianouma@cmu.edu}
\orcid{0009-0002-8634-130X}
\affiliation{%
  \institution{Carnegie Mellon University}
  \city{Pittsburgh}
  \state{PA}
  \country{USA}
  \postcode{15213}
}

\author{Kenneth Koedinger}
\email{koedinger@cmu.edu}
\affiliation{%
  \institution{Carnegie Mellon University}
  \city{Pittsburgh}
  \state{PA}
  \country{USA}
  \postcode{15213}
}

\author{Tongshuang Wu}
\email{sherryw@cs.cmu.edu}
\affiliation{%
  \institution{Carnegie Mellon University}
  \city{Pittsburgh}
  \state{PA}
  \country{USA}
  \postcode{15213}
}

\begin{abstract}
LLMs promise to democratize technical work in complex domains like programmatic data analysis, but not everyone benefits equally. We study how students with varied experiences use LLMs to complete Python-based data analysis in computational notebooks in a graduate course. Drawing on homework logs, recordings, and surveys from 36 students, we ask: Which experience matters most, and how does it shape AI use? Our mixed-methods analysis shows that technical experience -- not AI familiarity or communication skills -- remains a significant predictor of success. Students also vary widely in how they leverage LLMs, struggling at stages of forming intent, expressing inputs, interpreting outputs, and assessing results. We identify success and failure behaviors, such as providing context or decomposing prompts, that distinguish effective use. These findings inform AI literacy interventions, highlighting that lightweight demonstrations improve surface fluency but are insufficient; deeper training and scaffolds are needed to cultivate resilient AI use skills.
\end{abstract}



\maketitle


\section{Introduction}
\label{sec:intro}

Large language models (LLMs) hold the promise of democratizing complex technical tasks such as data analysis by supporting programming, analysis, and modeling~\cite{Ma2025-wf, Pickering2025-uj}.  
Despite the grand vision, evidence of who benefits and how remains mixed~\cite{paradis2025much}. Prior work reports conflicting effects of AI agents on programming efficiency~\cite{Becker2025-ho, song2024impact}; only some novices seem to perceive reduced barriers~\cite{Pickering2025-uj, Kazemitabaar2023-mj}.
In fact, performance often hinges on users' prior experience, ranging from domain knowledge to AI literacy~\cite{Chen2024-du, Nguyen2024-sd, zi2025would}. %
For instance, some work argues that less experienced developers benefit more from AI assistants~\cite{liang2024large, peng2023impact}, and others show that basic AI literacy helps users avoid common pitfalls in prompting or interpreting model outputs~\cite{puppart2025short,annapureddy2025generative}. 
However, these findings remain fragmented across different populations and tasks, \textbf{making it challenging to weigh which types of experience matter most} (e.g., whether we should focus on teaching domain knowledge or AI usage knowledge). 
Moreover, most prior analyses measure only outcomes (e.g., task success) and are conducted in controlled, short-term lab settings~\cite{Kazemitabaar2024-sb, Pickering2025-uj, Zheng2023-yx}. %
As a result, we have yet to demystify \textbf{how users' AI interaction behaviors get shaped by different experiences}, or the lack thereof, in real-world workflows. 

In this work, we \textbf{examine how varied forms of experience influence AI use and outcomes} in a naturalistic classroom setting (\cref{sec:class}): a graduate-level course, Data Science for Product Management (DSPM). The class included a diverse cohort of 36 students from CS majors to business students transitioning into technical roles, allowing us to examine varying domain expertise, LLM literacy, and communication skills. Students independently completed data analysis assignments (e.g., Data Cleaning) in Google Colab with full access to the built-in Gemini LLM assistant, while their workflows were captured through logs, surveys, and screen recordings. 
Through this setup, we investigate three questions with a mix of quantitative and qualitative methods (\cref{sec:method}):

\textbf{RQ1 (\cref{sec:skill-grade}): Do LLMs close the performance gap between students with different experience?}
We show that in open-ended, LLM-supported workflows, time constraints may remove the performance gap brought by prior technical experience. However, if given ample time, technical experience remains a significant predictor of success, and self-reported communication skills and LLM familiarity do not suffice.
This suggests that students with stronger programming and data science backgrounds outperform their peers when using AI in real-world tasks. 

\textbf{RQ2 (\cref{sec:ai-behavior}): How do experience differences translate to AI usage behaviors?}
Our analysis on over 7,000 log events shows that more technically experienced students use AI more strategically. They write clearer prompts and use AI proactively to improve plans (e.g., asking for better visualizations). In contrast, novices rely more on AI to resolve immediate challenges, such as debugging errors. 
These behavioral patterns offer \emph{a more nuanced explanation of why technical experience continues to matter} in AI-supported environments.

\textbf{RQ3 (\cref{sec:skill-to-teach}): What can we train students on AI usage?}
Comparing AI use behaviors before and after a lightweight instruction, we show that some AI-use skills (e.g., prompt quality) can be improved via live demonstration and extended time-on-task, while others (e.g., evaluation) likely require guided practice and feedback. 
We synthesize a set of AI-use competencies across \emph{Conceptual}, \emph{Procedural}, \emph{Metacognitive}, and \emph{Dispositional} knowledge dimensions.

Taken together, this work suggests that LLMs do not uniformly democratize programming but instead reshape when experience advantages appear. Our findings highlight AI literacy not as familiarity with tools, but as a set of transferable competencies for effective human–AI collaboration. We make three main contributions:

\begin{itemize}[leftmargin=1.2em,labelwidth=*,align=left]
    \item We provide \textit{empirical evidence} that \emph{LLMs do not fully close the experience gap} --- while AI may lower immediate barriers under time pressure, sustained success still depends on technical skills.
    
    \item We develop a \textit{behavior-oriented, LLM-based log annotation schema} that segments logs into semantically meaningful user stories, and we open-source our code: \url{https://github.com/mqo00/dspm}
    
    \item We identify \textit{concrete objectives and implications} for future curricula and tool design to train effective AI collaborators.
\end{itemize}

\section{Related Works}
\label{sec:related}

\subsection{Novice vs. Expert AI Programming Strategy}
\label{subsec:rw-expertise}
HCI and educational research highlight systematic differences between novices and experts when collaborating with AI for programming. 
In agent-based modeling, experts could benefit more than novices and rely on stronger skills in task decomposition and evaluation of AI outputs \cite{Chen2024-du}. 
In code comprehension, experts inspect the data more deeply, verify assumptions about transformations, and navigate outputs with clear goals --- behaviors that novices underuse \cite{Lum2025-sx}. Studies also document failure patterns such as ``prompting rabbit holes,'' where less-experienced users iterate on prompts without convergence \cite{Tie2024-ew}. 
{Beyond qualitative behavioral patterns, there has been mixed evidence regarding whether LLM use benefits people with various levels of expertise: some work found significant benefit from prior programming experience \cite{Kazemitabaar2023-mj}, some find no significance \cite{Pickering2025-uj}, while some found experts' performance using AI is actually hindered \cite{Becker2025-ho}.
Additionally, prior studies focus on challenges brought by domain expertise differences, commonly measured by the level of prior experience in programming via self-reported surveys \cite{Chen2024-du, Chen2024-gx, Lum2025-sx}, instead of analyzing different types of experiences.}
These findings motivate our work: measure whether LLMs reduce difference experience gaps, and analyze \emph{which behaviors} influence AI use efficacy in authentic workflows.

\subsection{Interfaces and Log Mining for AI-Assisted Data Analysis}
\label{subsec:rw-interfaces}

AI-based computational notebook assistants have investigated a vast design space of scaffolding users' interactions with LLMs during data analysis \cite{Mcnutt2023-po, Gu2024-lw}. 
For example, prior works have explored providing phasewise or stepwise decomposition interfaces \cite{Kazemitabaar2024-sb}, structuring how users express intent and context with ephemeral UI \cite{Cheng2024-pv}, manipulating intent \cite{Kim2025-bf}, and embedding explanations and documentation support \cite{Wang2022-zm}. These explorations brought different values to streamline the AI-assisted data analysis tasks, such as improving users' perceived control \cite{Kazemitabaar2024-sb} and lowering interaction barriers for novices \cite{Wang2022-zm, Cheng2024-pv}. However, they do not necessarily improve objective performance measures such as task success or completion time, as UI scaffolds on their own do not ensure learning outcomes or allow us to analyze the AI-use skills students need.

Beyond interface design, many efforts have explored mining logs of user interactions to derive useful insights. 
For example, recent works have explored logging and learning analytics in Jupyter notebooks to track students' interactions with a chatbot AI \cite{Valle-Torre2025-zi} or log novices' data science workflow \cite{Zhao2024-sk}.
These approaches highlight the promise of leveraging naturally occurring clickstreams to study novice reasoning at scale~\cite{claypool2001inferring, hijikata_implicit_2004, huang_improving_2012, chen2025need}, but also the challenges: raw logs are noisy, fragmented, and difficult to interpret without additional semantic structure, e.g., a long pause may indicate either confusion or deep focus.
To enable higher-level reasoning, we extract more \emph{semantically meaningful insights from the raw logs} by utilizing LLMs' reasoning and Theory-of-Mind capabilities, inspired by recent work~\cite{viswanathan2024large, rabinowitz2018machine, kosinski2024evaluating, strachan2024testing}.
We propose to combine log-based methods with behavior-grounded coding to identify success and failure patterns and identify needed AI skills. By semantically enriching logs into user story episodes, we aim to advance understanding of \emph{why} some learners succeed while others fail, and what competencies need to be taught (\cref{subsec:class-effect}). %

\begin{figure*}[th]
    \centering
    \includegraphics[width=.88\linewidth,trim={0 25 0 20},clip]{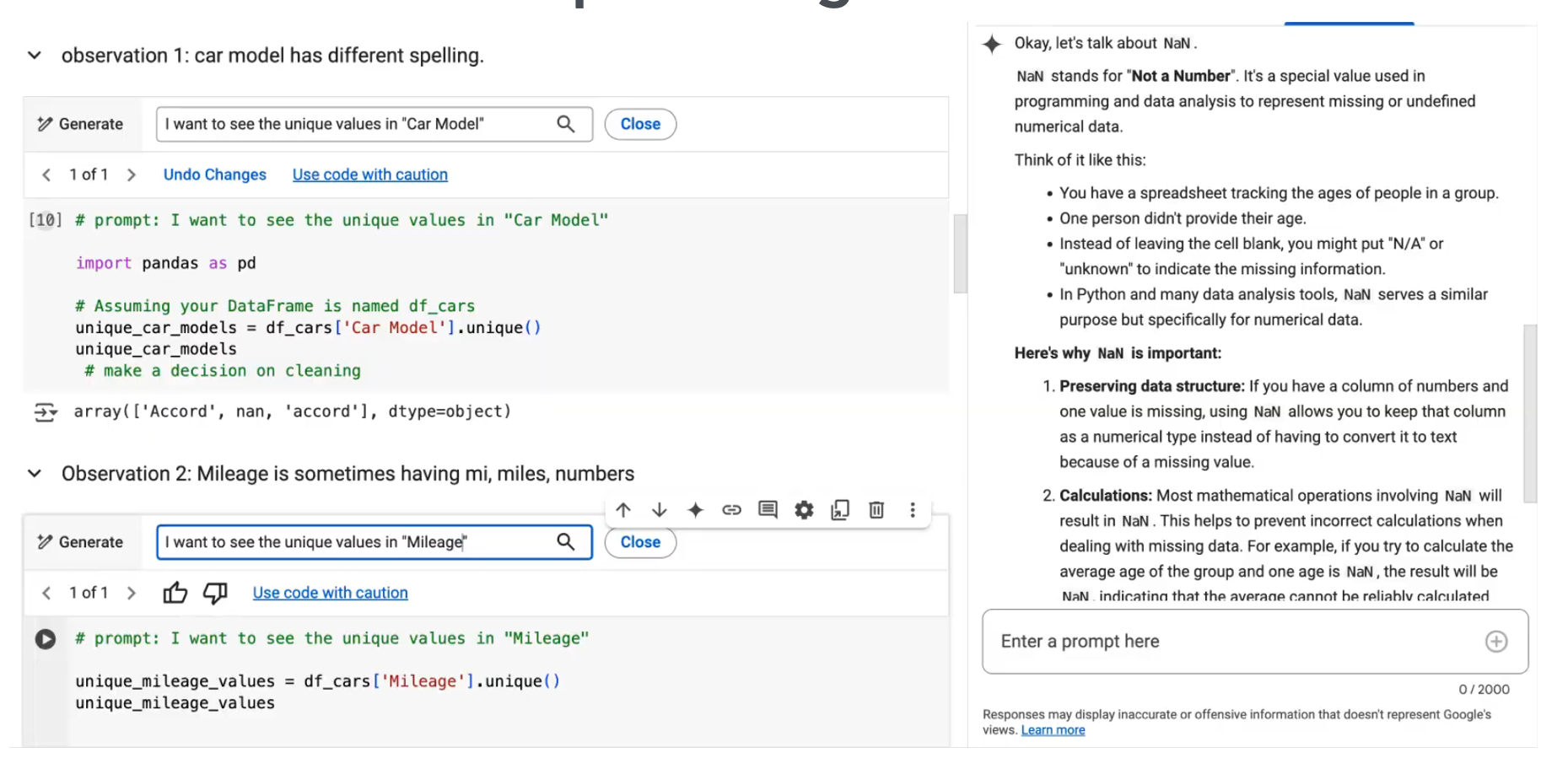}
    \caption{Google Colab notebook environment with embedded Gemini assistant. Students could interact with Gemini in two main ways: code cell generations (\texttt{generateCode}) and a conversational side chatbot (\texttt{converse}). There are also additional buttons that support error explanation, code explanation, and visualization generation.}
    \label{fig:colab-env}
    \Description{Screenshot of the Google Colab notebook environment showing the Gemini assistant. It displays code cells, a conversational sidebar, and buttons for explaining errors, explaining code, and generating visualizations.}
\end{figure*}

\subsection{Training for AI-Assisted Data Analysis and Effective LLM Use}
\label{subsec:rw-training}

Prior works have also explored instructions specifically for AI-assisted data analysis tasks. 
For example, \citet{Zheng2023-yx} allowed ChatGPT use in a graduate data science task and found that students reported that AI was more helpful for explaining code and less for open-ended problem-solving that needs context, especially on LLM-hard tasks where AI has limited capacities. LLM-enhanced feedback for an R-based data science tutoring system was also found to improve students' performance, while the verbosity of AI explanations sometimes caused cognitive overload \cite{Letteri2025-ga}. 
Nonetheless, these works mostly focus on analyzing surveys and prompt exchanges, leaving out behavioral log data that is important for capturing the full experiences and interactions with LLM \cite{Kazemitabaar2024-mg}. Moreover, these studies may not represent the most authentic workflows of how students would use LLM for computational data analysis, when there are popular tools such as Google Colab \cite{Google2025-pl} where AI is embedded in the notebook environment.

{Similar gaps exist in the broader LLM for CS education literature, where classroom-based studies, especially in graduate-level as in our context, remain limited \cite{Stone2024-oy, Raihan2025-fr}. In parallel, prior work primarily extracts student workflow or prompting patterns with surveys \cite{Gorson-Benario2025-ei, Scholl2024-jy}, interviews \cite{Tie2024-ew}, or think-aloud \cite{Lum2025-sx}, overlooking analysis on log data. Log data provides fine-grained traces of user behavior that reveal naturalistic actions without interference and helps us analyze interaction strategies linked to performance, which are not accessible through self-report, grades, or observation \cite{Dumais2014-dj, Ihantola2015-ei}.} 

There have also been some works that focus on task-agnostic, domain-general AI-use skills, such as prompt engineering. 
However, students do not learn from code feedback and self-experimentation \cite{Nguyen2024-sd}. Teaching specific prompt tips may not be very effective at influencing students' real success beyond perceived quality \cite{Sawalha2024-bz}, and risks getting obsolete with model updates \cite{Ma2025-wf}. Some prior works have shown promise in teaching requirement specification \cite{Ma2025-wf} or requirement elicitation \cite{Lojo2025-gh} for LLMs.
However, no work has comprehensively mapped AI-use competencies across the full workflow or disentangled which skills are harder to teach. %
Our study fills this gap by linking fine-grained behaviors to outcomes and quantifying post-demo shifts in skill acquisition.

\begin{table*}[htbp]
\centering
\caption{An overview and comparison of the homework setup, dataset, and rubrics. Overall task: analyze the data to gather some insights to present and guide product development across four stages of analysis (data cleaning, EDA, ML, storytelling).}

\footnotesize
\begin{tabular}{p{0.06\textwidth}  | p{0.43\textwidth}|p{0.44\textwidth}}
\toprule
& \textbf{HW0 Task} & \textbf{HW1-4 Task} \\
\midrule
Dataset 
    & Car dataset with 649 entries (crowdsourced from the prior semester of this class). Columns: Car Make, Model, Dealer/Individual, Price, Year, Location, Mileage, Doors, VIN\# {\color{cquote}(9 in total)}.
    & Game dataset with 1360 entries (adapted from a \href{https://www.kaggle.com/code/marianadeem755/gaming-evolution-exploring-videogames-1980-2023}{Kaggle dataset}, with renamed fields and inserted dirty data). Columns: name, description, developer, rating, genre, released\_on, \#total\_players, \#current\_players, \#saves, \#listed, \#reviews, selected\_reviews {\color{cquote}(12 in total)}. \\
\midrule
Cover Story
    & You're a product manager of Honda Accord, and the company is aiming to launch an online car price valuation tool to help buyers and sellers of used and new Honda Accord cars. Your team has collected a dataset for you by manually searching for 2008 to 2022 Honda Accords for sale in the U.S. 
    & 
    You're a product manager of Steam, a video game distribution platform, and the company is aiming to launch an analytics platform for game developers that provides insights into market trends. Your team has collected a dataset for you by manually scraping some popular games on Steam since 2000.  \\

\midrule
Template, \newline Timing \& \newline Tasks
    & \raisebox{-0.95\height}{\includegraphics[trim={0cm 2.5cm 0 0}, width=\linewidth,clip]{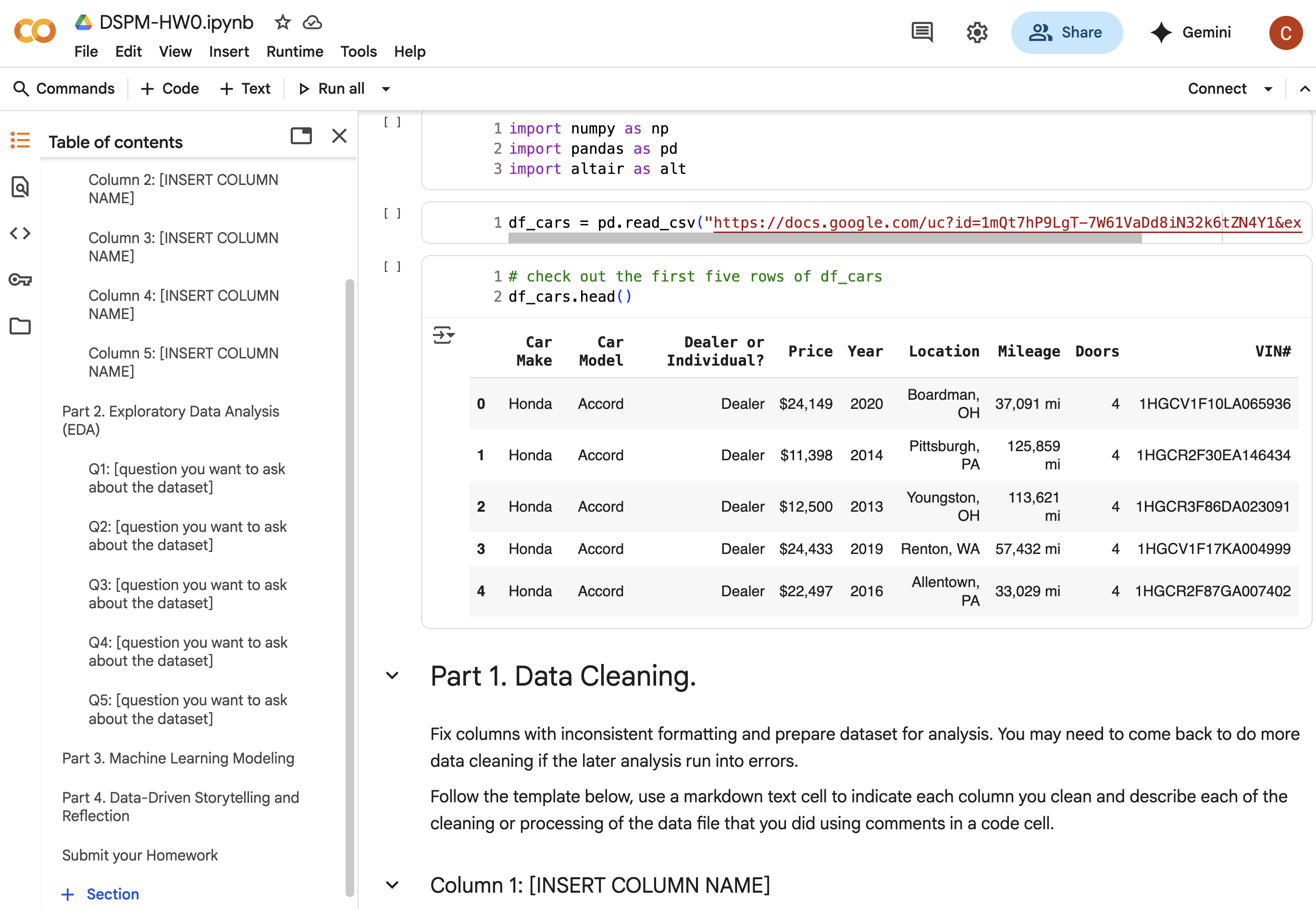} }
    & \raisebox{-0.95\height}{\includegraphics[trim={0cm 3cm 0 0}, width=\linewidth,clip]{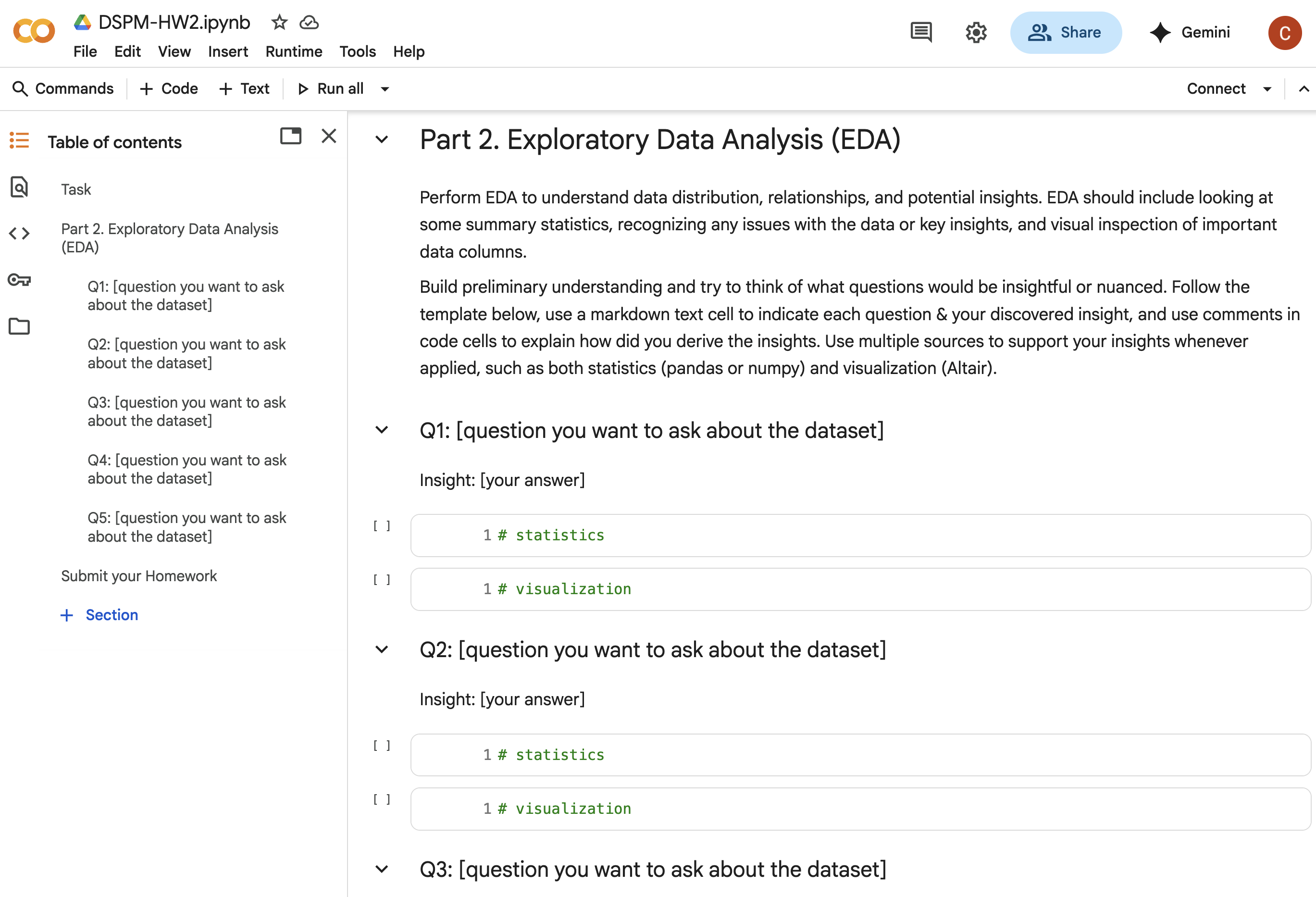}} \\
    & {\color{cquote}(screenshot of HW0 template for T1 data cleaning)}
    & {\color{cquote}(screenshot of HW2 template for T2 EDA)} \\
    & In-lecture activity. Four tasks (T1--T4), each $\sim$15 minutes. Half of the students use AI and half do not, counterbalanced across two tasks. 
    &  Bi-weekly assignments (HW1--HW4), each structured around T1--T4, with extended time for completion and free use of AI in Google Colab.  \\

\midrule
 Rubrics & 
 \multicolumn{2}{p{0.9\textwidth}}{
      \textbf{T1 Data Cleaning (max 15 pts):} Clean at least five columns of the dataset (e.g., duplicates, missing values, outliers, types, formats). 
      Rubric (for each of the 5 columns) — 0: no cleaning, 1: very incomplete, 2: partial, 3: complete. \newline
      \textbf{T2 EDA (max 45 pts):} Ask and answer meaningful data questions using stats/visualizations. 
      Rubric (for each of the 5 questions/answers/processes) — 0: missing, 1: simple/vague, 2: partial, 3: advanced/complete. \newline
      \textbf{T3 ML (max 15 pts):} Implement and document an ML workflow (train/test split, build model, measure performance, compare models, decide best model). 
      Rubric (for each of the 5 steps) — 0: not implemented, 1: very incomplete, 2: partial, 3: complete/correct. \newline
      \textbf{T4 Storytelling (max 15 pts):} Produce a clear data-driven report with framing, stats/modeling, visualization, storytelling, and data discussion. 
      Rubric (for each of the 5 aspects) — 0: missing, 1: very incomplete, 2: partial, 3: complete/clear. 
      }
  \\
\bottomrule
\end{tabular}
\label{table:hw-details}
\Description{Overview of homework setup, datasets, and grading rubrics. HW0 served as a diagnostic with short, timed tasks, while HW1–HW4 were bi-weekly assignments with extended time. Rubrics for each task (data cleaning, EDA, ML, storytelling) evaluated correctness and completeness on a 0–3 scale.}
\vspace{-10pt}
\end{table*}

\section{Course Structure and Task Setup}
\label{sec:class}

\subsection{Course Context}
\label{subsec:course}
Our IRB-approved study was conducted in Spring 2025 in \textit{Data Science for Product Managers} (DSPM) at an R1 University in the United States. The course serves students with highly diverse academic and professional backgrounds, including product managers, engineers, and graduates of non-technical disciplines. 
This diversity creates a pedagogical tension: assignments that are trivial for some are overwhelming for others.
The setting, therefore, provides an ideal opportunity to examine whether and how LLM support can reduce experience gaps in authentic workflows.

The class included four sequential assignments (HW1-HW4), each covering a primary data analysis stage: (1) data cleaning, (2) exploratory data analysis (EDA), (3) lightweight machine learning, and (4) data-driven storytelling. 
All assignments used the same dataset (more details in \cref{subsec:procedure}), 
enabling continuity across tasks. Lectures covered corresponding data science topics and basic GenAI use with a demo (more details in \cref{susbec:demo}).

Students worked in the Google Colab notebook with access to the built-in Gemini assistant \cite{Google2025-pl}\footnote{We chose Google Colab instead of Jupyter Notebook because it is easier to set up for students who do not have any programming background, and Colab has an in-house LLM (Gemini) that is free to use and provides various affordances. Specific model used by Google Colab in Spring 2025 is unknown, but it could be \href{https://cloud.google.com/vertex-ai/generative-ai/docs/learn/model-versions}{\texttt{gemini-2.0-flash} or \texttt{gemini-1.5-flash}}. We discuss the limitations of this approach compared to building on the open-source Jupyter Notebook in \cref{subsec:limitation}.}, but external GenAI systems such as ChatGPT were disallowed. Colab Gemini offered five interaction modes, including a chatbot, code cell suggestions, and buttons for error explanations, code explanations, and visualization generation with a dataframe (\cref{fig:colab-env} shows two of the main interfaces used for code generation: chatbot and code cell).

\subsection{Homework Design and Procedure}
\label{subsec:procedure}

Assignments were distributed as Google Colab notebook templates with minimal instructions and starter code, requiring students to ``fill in the blank'' using Python. We used unpopular datasets that are further mutated or collected via crowdsourcing in the prior semester, so solutions cannot be found via internet searches. We verified that the assignment could not be trivially completed by Colab Gemini or ChatGPT, by importing the template and dataset and prompting the models to complete all tasks with intuitive instructions.

Students first completed a \emph{diagnostic} HW0 during lecture to familiarize themselves with the tools and task sequence. HW0 covered all four tasks but used a different dataset than HW1-HW4 (described in \cref{table:hw-details}). We asked students to spend about 15 minutes on each task, where half the class completed the first two or last two tasks without using GenAI, providing a baseline for comparison. %
Students then completed HW1–HW4 over the semester on a bi-weekly basis, each focused on a different aforementioned task.
This setup enabled us to probe both short-term performance and longer-term workflow strategies, and to examine whether the LLM helps close the experience gap.
Details regarding the dataset, task, and rubrics are presented in \cref{table:hw-details}.

For each assignment, students submitted (1) a completed notebook, (2) exported log files of interactions, and (3) a post-survey reflecting on challenges and AI usage (details in \cref{subsec:data} and Appendix \ref{sec:appendix-surveys}). For HW0, students recorded their screen via Zoom, creating paired video–log traces for deep qualitative analysis. For HW1–HW4, bonus credit was offered for optional 60-minute think-alouds facilitated by one author, during which students would verbalize their thoughts while doing homework as usual.%

The assignments were graded by two teaching assistants (TAs) for the class. The TAs independently graded 20\% of the student submissions in HW0 and achieved a Cohen's kappa $\kappa=0.71$, which suggests substantial inter-rater reliability \cite{McHugh2012-sc}. Weekly calibration meetings were held to align rubric interpretation.

\subsection{Instructional Baseline Demo}
\label{susbec:demo}
Between HW0 and HW1, the instructor provided a 90-minute live-coding demonstration of the HW0 data cleaning task.
The demonstration showcased examples of AI use behaviors in the codebook (Table~\ref{tab:ai-codebook}), with explanations of when and why each behavior could be useful. For example, the instructor explained ``If you find unfamiliar concepts, you can actually ask Gemini to get you started. We can ask, \emph{What are some key steps in Data cleaning?}'' while typing and sending the actual question and walking through Gemini's answer.
The demo also covered the task environment (e.g., overview of Colab as a product) as well as the AI context space (e.g., the code cell generation has access to the entire notebook, while the chatbot sidebar does not have access to code cell outputs).
Students could re-watch the recording, too.
The demo's lecturing nature has limitations, as it can be less cognitively engaging than active learning and may create an illusion of fluency without fostering durable understanding \cite{Chi2014-ce, Deslauriers2019-ib}. However, this is a lightweight instruction baseline that enables analysis for what AI use competency needs to be taught with greater depth (RQ3). %

\section{Method}
\label{sec:method}

\label{subsec:analysis}

\begin{figure*}[tbp]
    \centering
    \includegraphics[trim=0 20 0 0, width=0.8\linewidth,clip]{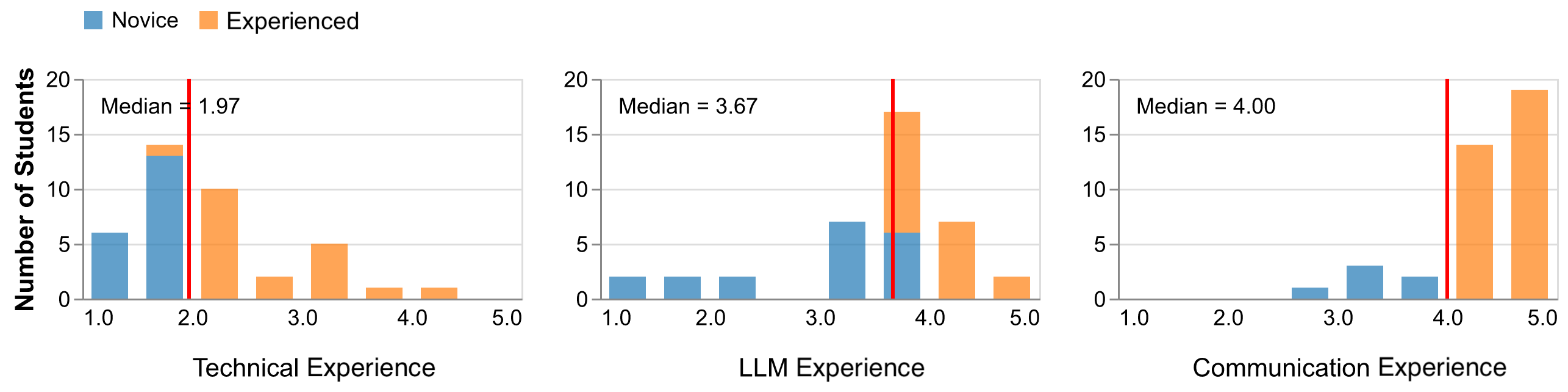}
\caption{Distribution of self-rated technical experience, LLM experience, and communication experience scores.}
\label{fig:expert-distr}
\Description{Three bar charts showing distributions of students’ self-rated technical experience, LLM experience, and communication experience. Technical and LLM experience vary widely, while communication ratings cluster on the higher end.}
\vspace{-10pt}
\end{figure*}

We used a mix of quantitative and qualitative methods to answer our research questions. %
\textbf{RQ1 (Do LLMs close the experience gap?)} was motivated by the premise that LLMs could level the field for less-experienced students, prompting us to statistically examine whether performance gaps persist and why. \textbf{RQ2 (How do differences in experience translate to AI usage behaviors?)} sought to understand students' real-world AI use patterns in behavioral logs and how these relate to experience levels and performance. \textbf{RQ3 (What can we train students on AI usage?)} was built on these findings to identify which AI competencies require deeper pedagogical investment to truly help students benefit from AI use.
Here, we provide a method overview, and some details (such as the regression factors and survey questions) are presented in the corresponding sections:
\begin{itemize}[leftmargin=1.3em]
    \item \textbf{RQ1
    } (\cref{sec:skill-grade}): 
    To assess whether different experience makes a significant difference in grades, 
    we conduct statistical testing, including non-parametric tests and regression on student homework data and experience scores;
    \item \textbf{RQ2
    } (\cref{sec:ai-behavior}): 
    To characterize students' AI usage behavior in relation to their task completion sequences, experience, and grades, 
    we perform thematic analysis to identify key AI behavior types, use them to annotate student logs (\cref{subsec:log-analysis}), and analyze the resulting distributions;
    \item \textbf{RQ3
    } (\cref{sec:skill-to-teach}): 
    To further identify skills essential for effective AI usage, we qualitatively analyzed survey responses as well as the aforementioned student log behaviors, identifying skill challenges and learning opportunities across conceptual, procedural, dispositional, and metacognitive dimensions. 
\end{itemize}

\subsection{Participants}
\label{subsec:participants}

A total of 41 students completed HW0, with a slight attrition to 36 students by HW4. Our analysis in this work was performed on these 36 students except for log analysis (\cref{subsec:log-analysis}), in which we filtered data to ensure quality (see details about sample size and data sources for each method in Appendix \ref{appen:method-detail}). Students came from a wide range of fields, including finance, telecommunications, journalism, physics, music, and computer science. This diversity highlights both the promise and the challenge of designing AI supports that work for ``everyone.''

{We used the widely adopted self-report practice and common questions in prior literature, such as ``How many years of experience do you have in using Python?'' \cite{Chen2024-du, Lum2025-sx, Chen2024-gx} in our surveys.} 
We collected pre-survey data on different self-reported experiences, including
\textit{implementation} (e.g., Python), \textit{domain} (e.g., data science), \textit{toolkit} (e.g., Colab/Jupyter Notebook, pandas, which are the specific environment and libraries used for our tasks), \textit{model} (e.g., GenAI/LLM proficiency), and \textit{non-technical skills} (e.g, communication abilities). 
We initially aimed to analyze whether these factors influence effective LLM use in different ways. However, in our class, some skills were moderately correlated (e.g., a Spearman correlation $\rho = 0.52$ for Python and data science items, and $\rho = 0.72$ between data science and toolkit items), making it difficult to disentangle their independent contributions (more on future work in \cref{subsec:limitation}).

Therefore, we operationalize students' \textbf{technical experience} by grouping the survey answers to programming, data science, and toolkit experience questions, such as year of experience using Python and ``How would you assess your level of expertise in [Exploratory data analysis]?''. Similarly, we merged highly correlated items into composite measures of \textbf{LLM experience} (e.g., ``How would you assess your level of expertise in [GenAI/LLM]?'') and \textbf{communication experience} (e.g., ``I am confident in my communication ability in general.'').
Items that were highly skewed toward zero or not informative were excluded, and we also used methods such as principal component analysis (PCA) \cite{Greenacre2022-rz} and variance inflation factors (VIF) \cite{Marcoulides2019-wz} to empirically verify our grouping (more details in Appendix \ref{sec:appendix-competency}). 
Specifically, multicolinearity is unlikely among the technical, LLM, and communication experience variables (Figure~\ref{fig:experience-overview}b), enabling further analysis of these skills. %
The distribution of these experiences among students is shown in \cref{fig:expert-distr}, where students were classified as novices or experienced using a median split of self-ratings.

\begin{figure*}[tbp]
    \centering
    \includegraphics[width=0.9\linewidth,trim={1cm 13.5cm 5.5cm 8.5cm},clip]{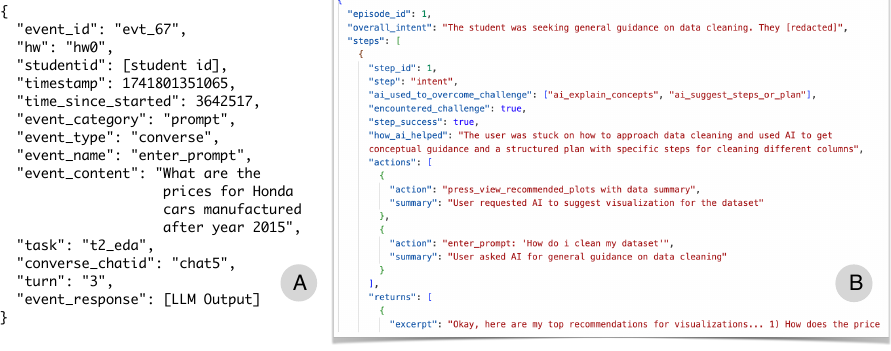}
    \caption{(A) Example raw log event format.
    (B) Example annotated log format. The student used AI in two \texttt{actions} (pressing a get visualizations button in Colab and entering a chat prompt ``How do I clean my dataset''), reflecting AI usage \code{ai\_suggest\_steps\_or\_plan} and \code{ai\_explain\_concepts} and indicating the \step{intent} of the first data cleaning episode.}
    \label{fig:ex-log-event}
    \Description{Example raw log event in JSON format showing metadata, prompt text, and LLM output, alongside an annotated version labeling AI usage behaviors and the inferred intent-formation step in a data-cleaning episode.}
\end{figure*}

\subsection{Data Collection}
\label{subsec:data}

Our dataset combines homework submissions, interaction logs, surveys, and screen recordings; for each student, we collected: 
\begin{enumerate}[labelwidth=*,leftmargin=1.3em,align=left]
    \item Notebooks and grades (HW0-HW4 submissions; rubric-based),
    \item Behavioral logs (HW0-HW4; code edits, executions, errors, prompts, model responses, timestamps, etc.), 
    \item Surveys (one pre-course survey, five post-HW surveys, one post-course survey) %
    \item Screen recordings (HW0 for all students; HW1–HW4 for think-aloud volunteers) %
\end{enumerate}

\paragraph{Surveys} We administered structured surveys at multiple points in the course (see Appendix \ref{sec:appendix-surveys} for full questions):

\begin{itemize}[leftmargin=1.3em]
  \item \textbf{Pre-Survey} (start of semester): Collected demographics, academic and professional backgrounds, Likert-scale questions on self-perceived experience, interest, confidence in personal vs. GenAI ability in different fields, etc. %
  
  \item \textbf{HW Post-Surveys} (after HW0-HW4): Captured task-specific experiences. Each survey asked which GenAI tools were used (e.g., Colab Gemini, ChatGPT), perceived task difficulty, effort, frustration, engagement, self-assessed success and confidence, and amount of peer help. Open-ended questions probed challenges with GenAI (and \emph{without} GenAI in HW0), how GenAI was used, its helpfulness, and what could improve performance.
  
  \item \textbf{Final Post-Survey} (end of semester): Collected self-assessed experience, interest, and confidence across domains (similar to pre-survey). Probed on similarities and differences across tasks, what they learned, and what they still wanted to learn.
\end{itemize}

\paragraph{Behavioral Logs}
We deployed a custom instrumentation in Colab to capture notebook interactions (code executions/errors), selected UI events (pressing explain-code/error buttons), and LLM use (prompts/responses). Students installed a browser script that intercepted requests to Colab's \texttt{converse} and \texttt{generateCode} endpoints and captured clickstream events. The script added a ``Download History'' button to the Colab toolbar and stored code cells, chat messages, event metadata, and responses to browser localStorage, which students exported as JSON files upon submission (or auto-save when the storage is full). %
In total, we collected 16,315 logged events (e.g., \cref{fig:ex-log-event}A) and 44 hours of screen recordings across five assignments.
Logs captured a rich sequence of AI interactions, code edits, and executions, supporting both quantitative distributional analyses and qualitative episode-based coding (\cref{subsec:log-analysis}).

\subsection{LLM-assisted Log Data Annotation} 
\label{subsec:log-analysis}

\begin{table*}[t]
\centering
\fontsize{8}{9}\selectfont
\caption{Codebook for AI usage behaviors. 
{
Code reflects different \textbf{conceptual} \emph{(\textbf{what} to use AI for)} and \textbf{procedural} (\emph{\textbf{how} to use AI effectively through prompting)} knowledge dimensions.
}
For the \emph{prompt (quality)} category, we provide \emph{negative} examples since other rows already demonstrate positive examples.
}
\label{tab:ai-codebook}

\vspace{-10pt}
\renewcommand{\arraystretch}{1.3}

\begin{tabular}{c|c|l|p{0.55\textwidth}}
\toprule
 \textbf{Dim.} & \textbf{Category} & \textbf{Code} & \textbf{Definition and Examples} \\
\midrule
\setlength{\tabcolsep}{3pt}

\multirow{9}{*}{\rotatebox[origin=c]{90}{\textbf{Conceptual} Knowledge Dimension}} 
 & \multirow{3}{*}{\rotatebox[origin=c]{90}{Improve (Plan)}} 
    & \texttt{ai\_breakdown\_intent} 
    & Decompose a complex goal into smaller pieces. 
      E.g., ``How to improve performance of model. give step by step.'' \\

 & 
    & \texttt{ai\_improve\_prompt} 
    & Refine the wording of a prompt for clarity and specificity. 
      E.g., ``plz improve my prompt: [prompt]'' \\

 & 
    & \texttt{ai\_suggest\_steps\_or\_plan} 
    & Provide a step-by-step workflow (e.g., data cleaning → modeling). 
      E.g., ``How to do data cleaning?'' or ``How to build ML models?'' \\
\cmidrule{2-4}

 & \multirow{2}{*}{\rotatebox[origin=c]{90}{Code}} 
    & \texttt{ai\_generate\_code} 
    & Produce code that implements a requested action. 
      E.g., ``Write code to calculate the distribution of cars by location.'' \\

 &
    & \texttt{ai\_edit\_partial\_code} 
    & Edit a specific snippet or function rather than rewriting the whole cell. \\
\cmidrule{2-4}

 & \multirow{3}{*}{\rotatebox[origin=c]{90}{Explain}} 
    & \texttt{ai\_explain\_bug\_or\_error} 
    & Explain an error or traceback and outline a concrete fix. E.g., Copy-paste error message and use "explain error" to understand and fix it.\\
    
 &
    & \texttt{ai\_explain\_code\_or\_api} 
    & Interpret code or explain APIs, such as describing what a function does and how to use it. E.g., ``Explain what pd.get\_dummies does.'' \\

 &
    & \texttt{ai\_explain\_concepts} 
    & Provide explanations of concepts. E.g., ``How do data cleaning steps work?''\\
\cmidrule{2-4}

 & \multirow{1}{*}{\rotatebox[origin=l]{90}{Eval}} 
    & \texttt{ai\_critique\_output} 
    & Evaluate results or insights for correctness, alignment with goals, and suggest improvements. E.g., ``Check my insight and provide improvements'', ``Is it a good model to work with?'' \\
\midrule

\multirow{3}{*}{\rotatebox[origin=c]{90}{\textbf{Procedural}}} 
   & \multirow{3}{*}{\rotatebox[origin=c]{90}{Prompt (Quality)}} 
    & \texttt{clear\_instruction\_in\_prompt} 
    & Write prompts with explicit actions and desired outcomes. 
      Example of a poor prompt: “Fix it’’ without identifying what “it’’ is. \\

 &
    & \texttt{decompose\_task\_in\_prompt} 
    & State the problem as ordered sub-tasks or ``step-by-step'' instructions within the prompt. E.g., ``This was my result array [copied output]'' without intent would be too vague.\\
    
 &
    & \texttt{provide\_context\_in\_prompt} %
    & Include necessary details (e.g., dataframe names, prior steps) and avoid vague pronouns. E.g., ``Don't make up non-existent column name'' without providing actual context does not help resolve hallucination.\\

\bottomrule
\end{tabular}
\Description{Codebook listing AI usage behaviors across conceptual and procedural knowledge dimensions with definitions and examples, including planning, coding, explanation, evaluation, and prompt-quality categories.}
\end{table*}

\label{subsubsec:qual-log-analysis}

To answer how students use LLMs (RQ2) and what we should teach on effective LLM usage (RQ3), we need to identify the core \textbf{AI usage behaviors}. To this end, we analyze students' log data to develop a codebook to capture behavioral patterns in authentic problem-solving processes.

\paragraph{Log Parsing}
Raw logs consist of fragmented sequences of actions that lack meaningful semantics. To recover actual student usage and rationale, we analyze user logs through a hierarchy of \texttt{episodes}, \texttt{steps}, and \texttt{log actions} (as shown in \cref{fig:ex-log-event}B).
Specifically, we segmented logs into \textbf{\texttt{episodes}} --- goal-driven units of activity (e.g., "remove duplicates in data column X"). 
Inspired by Norman's Stages of Action for user cognitive processes~\cite{tenner2015design}, we further broke each episode into four \textbf{\texttt{steps}} --- forming \step{intent}, expressing the intent as an \step{input} (e.g., writing code or prompt), obtaining and \step{understanding output} (e.g. inspecting the code execution results or errors), and \step{assessing} output (e.g., deciding if the intent is sufficiently fulfilled). Each step contains a sequence of \textbf{\texttt{log actions}}. Within each step, we annotated whether the student encountered \emph{challenges}, whether the step \emph{succeeded or failed}, what \emph{AI usage behavior} occurred, and what additional AI usage behavior \emph{could have helped}, or missed opportunities of benefiting from AI usage. %

\paragraph{Codebook for AI usage behavior analysis.} 
To develop the codebook (Table~\ref{tab:ai-codebook}), two authors thematically analyzed a subset of homework logs and screen recordings, following a hybrid of deductive and inductive thematic analysis approach \cite{braun2006using, fereday2006demonstrating}.
To maximize verifiability, we used five logs and screen recordings from HW0 and five logs and think-aloud recordings from HW1-4.
The AI code types in our codebook are characterized by two dimensions: \emph{conceptual knowledge} on what to use AI for (using AI to improve plans, code, explain concepts, or evaluate deliverables), and \emph{procedural knowledge} of how to use AI via good quality prompts, as using an AI is not the same as using the AI \emph{well}.

Our codebook integrates and extends prior work in multiple layers --- it links human-AI collaboration with Bloom's knowledge dimensions \cite{Krathwohl2002-nu}, which enables pedagogical alignment (more in \cref{sec:skill-to-teach}). %
The knowledge \textit{dimensions} offer a complementary perspective to the current prompt coding schema based on Bloom's cognitive process \cite{Chen2024-gx}. 
The code \textit{categories} build on and decompose existing themes that characterize human-AI collaborative behaviors, including planning, evaluating, and coding \cite{Chen2024-du}. Our prompting \textit{codes} also echo existing strategies such as ``be specific,'' ``write accurate requirement'' (\code{clear\_instruction\_in\_prompt}) and ``set the scene'' (\code{provide\_context\_in\_prompt}) \cite{Tassoti2024-tl, Ma2025-wf}.

\paragraph{Scale up with automated log annotation}
Because manual annotation was time-intensive\footnote{Even with the video and with one author already familiar with the session, on average, it took > two hours to go through each log. 
It is even more difficult to decipher logs that do not come with video recordings, due to, e.g., readability issues (certain log events would contain more notebook context information with code cell metadata that are hard to manually read through).}, we scaled up by using an LLM to label log steps with the codebook (\texttt{claude-sonnet-4-20250514} for its reasoning capability, with temperature=0 for deterministic generations, full prompts in \cref{sec:appendix-prompt}).
We had the model identify the challenges as well as AI usage --- either the student used AI ($used$), or the student missed an AI use opportunity ($missed$)\footnote{Annotation output contains other features to support the model's inference time reasoning and human verification, e.g., \texttt{id}, reasoning behind certain label, etc. 
Complete prompt with the output format and hyperparameters are in Appendix \ref{sec:appendix-prompt}.}.

\paragraph{Data filtering. } 
{
To ensure the reliability of our quantitative log analyses that examine experience effects and pre-post instructional changes on interaction behaviors, we applied a series of filters to improve data quality.
}
Specifically, we filtered out\footnote{We required students to use the built-in Gemini and avoid using other LLMs and added the precaution of asking students to self-report their additional LLM model use, if any. However, we still noticed students using external AI tools, which tend to result in reasonable grades but no or few user logs captured, or logs with abnormal behaviors, e.g., copy-pasted large blocks of code with unfounded originality.}
students who self-reported other tool use like ChatGPT or displayed other tool use in recording (if available), as well as those who disengaged or spent too much time working on tasks without AI, reflected as log length being less than the median. 
{
To enable comparisons of behavioral changes before and after instructions, we only apply the quantitative analysis over logs on T1 data cleaning and T2 EDA, as most interactions come from tasks 1 and 2 in HW0 due to time limits.
}
As a result, we only annotated 7,315 logged events from 28 students across HW0-2.\footnote{Each homework may include a different subset of student submissions after filtering. We conduct log annotations over a total of 36 submissions (20 from experienced, 16 from novices). Note that ``experienced'' vs. ``novice'' denote technical experience here and for the rest of the paper.}

\paragraph{Validation of LLM annotation. }

Our annotation produced, on average, 7.6 episodes per log file, with 5.6 steps per episode.
Out of a total of 7315 log events, there are 1483 steps (\step{form intent}, \step{write code}, \step{understand output}, and \step{assess output}) across all episodes, and 920 (62.0\%) of them are annotated with AI usage ($used + missed$). The distribution of average AI usage frequencies by all the codes is shown in \cref{fig:code-distr-step}. 
An author verified {10\% of} the log annotations, and LLM achieved an accuracy of 75\% {(specifically, 81.3\% accuracy for \textit{step success}, 76.7\% for \textit{encountered challenge}, and 70.9\% for \textit{AI use behaviors} annotations)}. %

\section{RQ1: Do LLMs close experience gap?} 
\label{sec:skill-grade}

Students engaged substantially with AI during coursework: 15\% of logged events were prompt-related, and homework post-surveys all confirmed that students spent non-trivial amounts of time with AI tools. Despite substantial AI usage, we found that technically experienced students seem to benefit more from LLM use than novices if given ample time.

We hypothesized that students with higher technical experience would perform better in our programmatic data science tasks. To examine if using LLM helps bridge this performance gap, we compared performance between more technically experienced students and novices in the timed HW0 under conditions with and without LLMs (\cref{table:hw-details}). 
A non-parametric Mann–Whitney U test (as data violated normality assumptions) \cite{McKnight2010-nr} showed a significant technical experience gap when students did not use LLMs ($p = 0.027 < 0.05$), but the gap disappeared when LLM was used ($p > 0.05$)\footnote{We also verified that the communication and LLM experience do not lead to any significant performance gap with or without LLM in HW0.}. This suggests that \textbf{under time pressure, LLMs may offset technical experience's advantages}.

\begin{table}[]
\centering
\small
\caption{Linear mixed-effects regression predicting homework performance from different experiences. Participant and homework were used as random intercepts.}
\begin{tabular}{r|rrrr}
\toprule
Predictor & Estimate & SE & df & p \\
\midrule
Intercept & 89.88 & 17.84 & 33.82 & $<.001^{***}$ \\
\textbf{Technical} & 6.09 & 2.86 & 30.40 & $.041^{*}$ \\
LLM & $-3.05$ & 2.69 & 30.77 & .266 \\
Commun. & $-2.54$ & 2.85 & 30.97 & .381 \\
\bottomrule
\end{tabular}
\Description{Regression results table showing that technical experience significantly predicts homework grades, whereas LLM experience and communication skills do not.}
\label{tab:lmer-results}
\end{table}

\label{subsec:method-lmer}

To further probe the interplay between different experiences' influence over LLM use in more authentic scenarios without the time constraints, we assess whether different experiences make a significant difference in grades in HW1–HW4.

We fit a linear mixed-effects regression model \cite{Bates2015-fl} with grades (normalized scores) as the outcome and technical, LLM, and communication experience scores as predictors, including random intercepts for participant and assignment. 
As shown in (\cref{tab:lmer-results}), only \textit{technical experience} significantly predicted higher grades ($\beta = 6.09, p = .041$). LLM and communication experience had no significant effects and did not predict grades. This suggests that even with equal access to the Colab Gemini assistant, students with stronger programming and data science backgrounds continued to outperform peers when given ample time. In other words, \textbf{the technical experience gap may not be closed by an LLM in a more realistic setup}, which is consistent with some prior literature \cite{Chen2024-du}.

\label{subsec:result-lmer}

Note that students in our class had a median of 4/5 for their communication experience (see \cref{fig:expert-distr}), which means that the communication factor may not have a lot of variance to predict grades. Meanwhile, students' median of LLM experience is 3.67/5, but everyday use of LLMs or their self-rated LLM knowledge may not straightforwardly transfer to complex programming data science tasks. 
{Our HW0 tasks without LLM use may serve as a proxy of pre-assessment over technical experience, as partially validated by the significant performance difference between more technically experienced and novice students}. Nevertheless, future work should explore more rigorous ways to measure experience and recruit a more diverse set of students (more in \cref{subsec:limitation}).

\section{RQ2: How do experience differences translate to AI usage behaviors?} %
\label{sec:ai-behavior}

The analysis in RQ1 showed that students' self-perceived technical experience continued to predict performance even when all students had access to the Colab Gemini assistant. Here, we dive into student log data to examine why: \emph{how students use AI during workflows, and how it relates to different levels of technical experience?}

\begin{figure}
\includegraphics[trim=0 25cm 40cm 0cm, width=1\linewidth,clip]{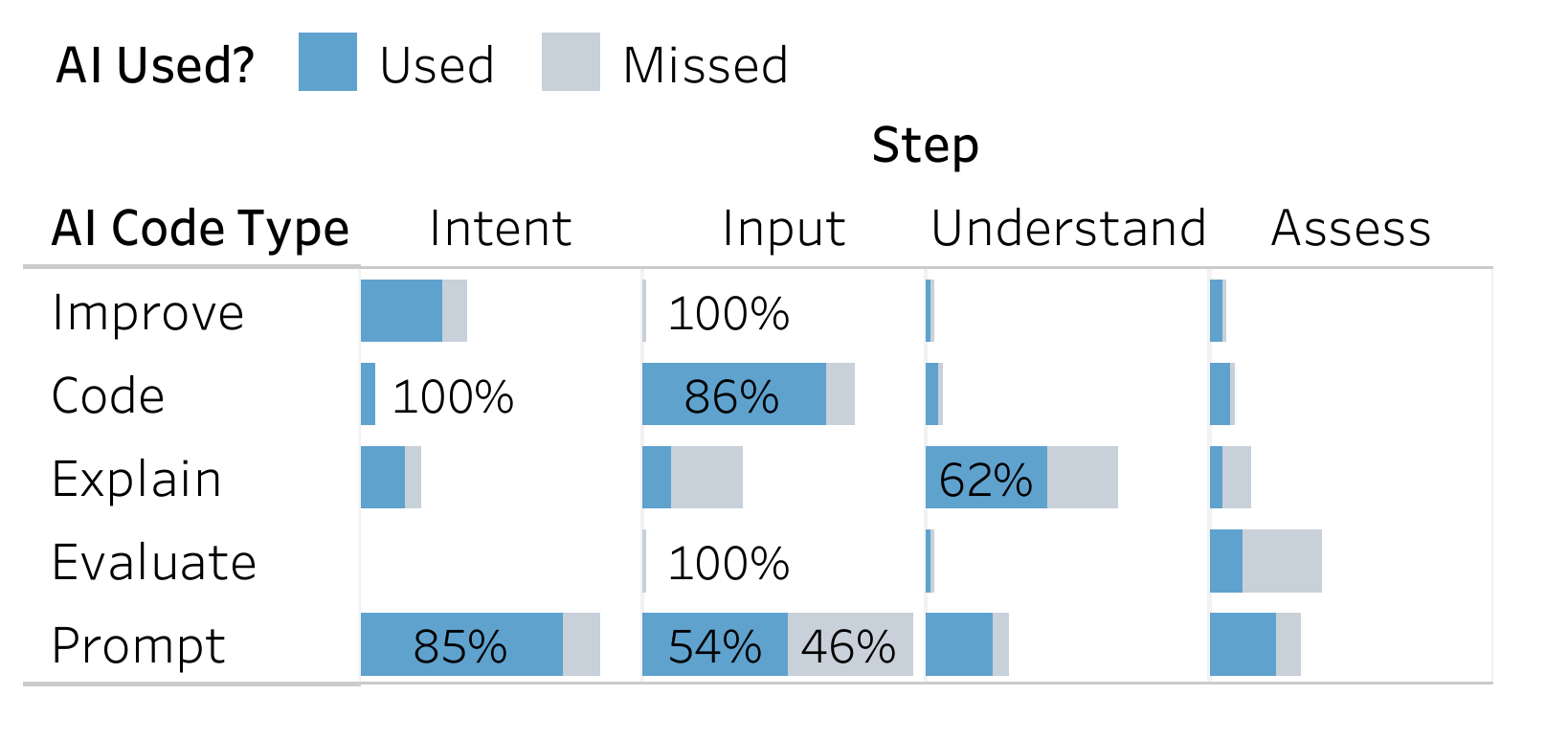}
\vspace{-10pt}
 \caption{Distribution of the AI usage (both $used$ and $missed$) aggregated by code type and episode steps. Bar length represents the number of steps involving this AI use, while the percentages denote the proportion of $used$ vs. $missed$ opportunities. 100\% means that all AI usage is $used$ (if blue) or $missed$ opportunities (if grey).}
\label{fig:ai-behavior-step}
\Description{Stacked bar chart showing AI usage across four episode stages—intent, input, understand, assess—with bars indicating used versus missed opportunities for planning, coding, explanation, evaluation, and prompt-quality behaviors.}
\end{figure}

\paragraphBold{Steps vs. AI usage: AI can help throughout stages, with clear mapping of ``how''.}
There is a clear relation between how AI is used and when (\cref{fig:ai-behavior-step}). 
Behaviors associated with \texttt{improving} (e.g., \code{ai\_suggest\_steps\_or\_plan}) occur most frequently when students try to form \step{intent}. We see similar trends for \texttt{AI coding} during \step{input}, \texttt{AI explaining} during \step{understand}, and \texttt{AI evaluating} during \step{assess} step.

Looking more into the ratio of AI usage, we notice \textbf{students often missed opportunities to use AI for explanation and evaluation}, despite \texttt{explain} code and \texttt{explain} errors being supported by one-click features in Colab notebook.
As discussed in \cref{subsec:class-effect}, we suspect students felt overloaded by AI output.

We also observe that \texttt{prompt quality} could have been largely improved at  \step{input} step (46\% missed opportunities).
Instead of clearly instructing the AI (e.g., “write the code to calculate the distribution of the cars by location”) and adding appropriate context (e.g., “help clean up \code{df\_games["\#total\_players"]}”), some students wrote vague prompts such as “fix it” or pasted raw outputs without clear intent in prompt. 
Without \code{clear\_instruction\_in\_prompt} or \code{provide\_context\_in\_prompt}, Gemini often produced irrelevant or hallucinated code.
This echoes prior work that highlights \textbf{the importance of teaching students to express their requirements to LLMs}~\cite{Ma2025-wf}.

\paragraphBold{Experience vs. AI usage: More technically experienced students could use AI more strategically, and have more cognitive budget to explore.} 

\begin{figure*}[!h]
    \centering
    \vspace{-10pt}
    \includegraphics[width=0.7\linewidth,trim={0 24cm 22cm 0},clip]{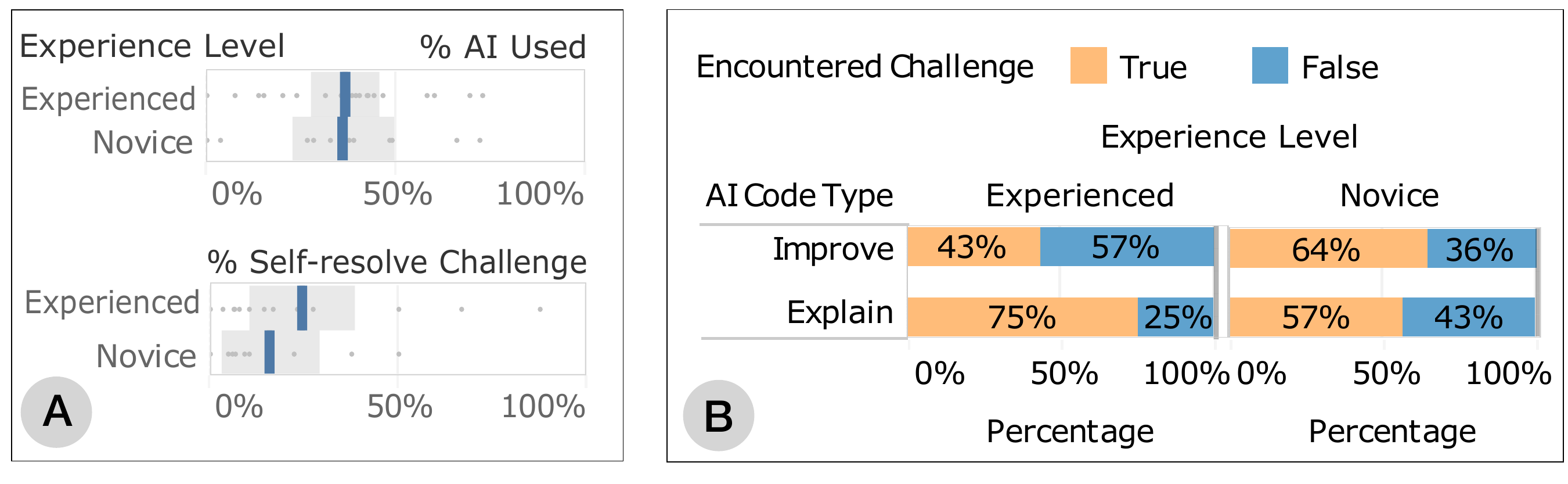}
    \vspace{-5pt}
    \caption{(A) While experienced students and novices used AI with similar frequencies, experienced students might be more able to resolve challenges without using AI. 
    (B) Experienced students might strategically distribute effort to ask AI to \texttt{plan} and \texttt{improve} or \texttt{explain}, different from novices.} 
    \label{fig:expertise_strategy}
    \Description{Two sets of bar charts comparing novices and experienced students. One shows that experienced students resolve more challenges without AI; the other shows differences in how each group requests planning, explanation, and coding support from AI.}
    \vspace{-10pt}
\end{figure*}

\begin{figure*}[!htbp]
    \centering
    \vspace{-10pt}
    \includegraphics[width=0.7\linewidth,trim={0cm 12.8cm 10cm 2cm},clip]{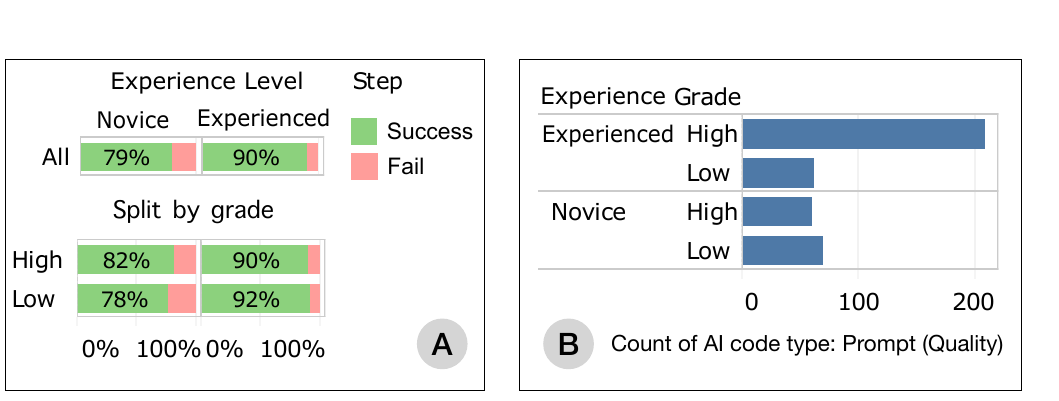}
    \vspace{-5pt}
    \caption{(A) Experienced students achieve a higher step success rate than novices when using AI to assist with \texttt{coding} (\code{ai\_generate\_code} and \code{ai\_edit\_partial\_code}); Dividing by grades shows more variations.
    (B) Experienced students with high grades contributed to most prompts with high prompt (quality) attributes. High or low grades are split based on the median.} 
    \label{fig:expertise-to-code}
    \Description{Bar charts showing experienced students' higher success rates in AI-assisted code generation and more frequent use of high-quality prompt attributes compared to novices and lower-performing students.}
\end{figure*}

As shown in \cref{fig:expertise_strategy}A, experienced students and novices used AI at similar frequencies (in around 36\% of steps). However, \emph{experienced students might be less pressed into using AI when facing challenges} --- they were more capable of resolving challenges without using AI than novices (\cref{fig:expertise_strategy}A). %
Nevertheless, when experienced students did use AI (\cref{fig:expertise_strategy}B), they might ask AI for \texttt{explanation} more often than novices when they ran into challenges (75\% > 57\%). Meanwhile, their requests for AI to \texttt{improve} or \texttt{plan} seemed less tied to challenges than novices (43\% < 64\%). %
This different AI use strategy may reflect technically experienced students' clearer awareness of when and what type of help is needed. They may be better able to identify the challenges and seek targeted clarification and better request planning support proactively before encountering challenges, compared to novices.

Moreover, our logs reveal patterns echoing prior research~\cite{guo2024personalized, palta2025speaking}, such as experts are better at generating code using AI and prompting AI, which could result from having more domain knowledge. As shown in \cref{fig:expertise-to-code}A,
\emph{technically experienced students generate code with AI more successfully} (90\% > 79\%).
Further analyzing with student grades, it seemed like novices might rely more on AI coding being accurate to achieve higher scores --- novices with higher grades also had a higher AI coding success rate than novices with lower scores (82\%> 78\%). In contrast, experienced students achieving high scores actually had slightly lower success on AI coding steps (90\% < 92\%).
As shown in \cref{fig:expertise-to-code}B, \emph{technically experienced students better describe their needs}. Prompts from experienced students who achieved high scores included twice as many quality attributes of \code{clear\_instruction}, \code{decomposed\_task}, and \code{provide\_context}. %

We believe these experience gaps might be closed \emph{by training on \textbf{AI use skills}}.
For example, a sequence of \code{ai\_breakdown\_intent} $\rightarrow$ \code{ai\_generate\_code} might help a novice obtain more easily digestible code; if followed by ``further simplify your explanation'', their chance of correctly understanding AI code can be improved. %
Similarly, students can ask for \code{ai\_suggest\_steps\_or\_plans} to familiarize themselves with the task, or \code{ai\_explain\_concepts} to understand domain-specific keywords.
However, our analysis shows that much of the student grade and effective AI usage still comes down to technical experience. 
In the next section, we explore what we should we do better in teaching \emph{AI use skills}. 

\section{RQ3: What can we train students on AI usage?} 
\label{sec:skill-to-teach}

The findings in RQ2 suggest that technically experienced students use AI more strategically and effectively than novices. Here, we ask: \emph{What are some skills and how to teach better AI users?} %

By comparing student behaviors pre- to post-demo (\cref{susbec:demo}), we collected preliminary signals on what AI use skills might be more easily acquirable than others.
Note that in addition to instructions on AI use (and regular lectures on programming and data science knowledge), students moved from HW0's 15-minute tasks to bi-weekly assignments (\cref{table:hw-details}).
Therefore, any improvement pre- to post-demo is a result of both extended time-on-task and the instruction, and any room for improvement builds on top of gains already afforded by extended practice time and baseline instruction.

\begin{table*}[tb]
\centering
\small
\caption{The codebook of identified AI use skills by knowledge dimension.}
\label{tab:ai-use-los}
\vspace{-5pt}
\begin{tabular}{p{0.14\linewidth}p{0.82\linewidth}}
\toprule
\textbf{Knowledge Dim.} & \textbf{AI Use Learning Objective Description} \\
\midrule
\textbf{[Meta]cognitive} & \emph{Monitor when to delegate things to AI and decide what to delegate.} \\ \midrule
\textbf{[Conc]eptual} & \emph{Knowing what can be delegated to AI (task space, context space) with calibrated expectations of AI capacity (not under- or over-estimate).} \\ \midrule
{[Conc-Task]} & Knowing AI Task Space: \code{plan}, \code{critique}, \code{improve prompt}, \code{decompose}, \code{explain}, \code{edit partial code}, etc. \\ \midrule
{[Conc-Context]} & Knowing AI Context Space: base Colab Gemini shows no access to closed chats, unreferenced code, imported dataset; different UI (generateCode vs. converse) has different access to context, by the time we ran the study. \\ \midrule
\textbf{[Proc]edural} & \emph{Knowing how to best delegate things to AI for instruction following (clear instruction, context, broken down if overly complex).} \\ \midrule
{[Proc-Clear]} & Give \code{clear instructions} with action verb to AI. \\ \midrule
{[Proc-Context]} & Provide \code{correct context} to AI without vague pronouns. \\ \midrule
{[Proc-Decompose]} & Provide an achievable, \code{decomposed goal} for AI instead of asking AI to do too many things at the same time. \\ \midrule
\textbf{[Disp]ositional} & \emph{Willing to actively use AI with trial and error, read and interpret long outputs, and persist through challenging tasks with high intrinsic cognitive load.} \\
\bottomrule
\end{tabular}
\Description{Learning objectives for AI-use skills grouped by cognitive, conceptual, procedural, and dispositional dimensions, describing what students should know and how they should engage when using AI tools.}
\end{table*}

\begin{figure}
    \centering
    \includegraphics[width=.75\linewidth,trim={0 0 0 0},clip]{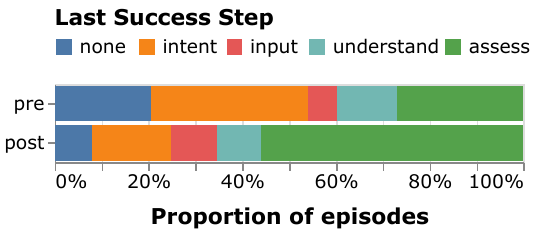}
    \caption{The distribution of the last successful step in an episode, before and after class demonstration. More students post-instructions \emph{can} successfully get to last step of \step{assess} in their workflow comparing to HW0.}
    \label{fig:last-success-step}
    \vspace{-10pt}
    \Description{Bar chart showing distribution of the last successful step in each analysis episode before and after instruction. Post-instruction episodes progress further, especially to the assessment stage.}
\end{figure}

\begin{figure*}[tb]
\centering
    \centering
    \includegraphics[trim=0 0 0 0, width=\linewidth]{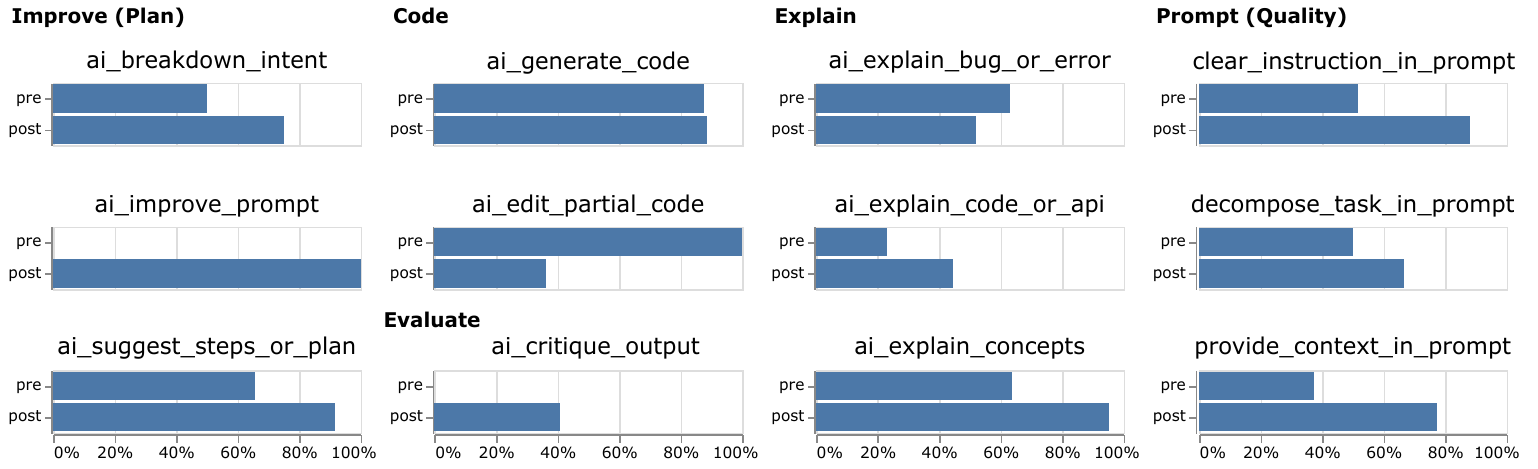}
    \caption{Appropriate AI use ratio pre- to post-instruction. 
    Ratio = $used / (used + missed)$.}
    \label{fig:ai_use_ratio_code_distr}
    \Description{Bar charts comparing pre- and post-instruction appropriate AI use ratios across improvement, coding, explanation, evaluation, and prompt-quality behaviors. Prompt-quality behaviors show the largest gains; evaluation remains limited.}
\end{figure*}

\paragraphBold{In-class demonstration and extended time-on-task improved AI use success, especially by improving prompt, but evaluation behaviors need more targeted instructions.} 
\label{subsec:class-effect}

As shown in \cref{fig:last-success-step}, students were able to sustain progress further along the problem-solving process at post-demo. 
In HW0, many episodes stalled at early stages (examples of student-reported challenges in \cref{tab:task-challenges}; Appendix \ref{sec:appendix-challenge}) or ended without any successful step (already failed at \step{intent}, denoted by “none”). After being provided with baseline instructions and ample time, more episodes advanced to later steps in the sequence of \step{intent} $\rightarrow$ \step{input} $\rightarrow$ \step{understand} $\rightarrow$ \step{assess} (as described in \cref{subsec:log-analysis}).

We also computed an appropriate AI use ratio ($used / (used + missed)$) to measure whether students capitalized on available AI opportunities.
As shown in \cref{fig:ai_use_ratio_code_distr}, appropriate AI use ratios increased post-demo in many behaviors.
While not fully mastered or statistically compared, there is a notable \textbf{progress in \code{prompt quality}} behaviors by an average of 30\% in students' AI use ratio. 
Students also showed modest gains in asking \code{ai\_explain\_concepts}. 
However, \textbf{\code{evaluation} behaviors remained a bottleneck} (e.g., post-demo ratios of using AI to critique, asking for code or bug explanations are all < 50\%). This suggests a shared hesitation to request or process complex evaluative information and use AI to deepen understanding. These observations hint towards under-taught AI use skills, such as the disposition to persist through trial-and-error, read and interpret long outputs, and refine AI's outputs. %

Note that some AI use behaviors like \code{provide\_context\_in\_prompt} are much more frequent (18.06\%) than others. The appropriate AI use ratios can be less interpretable and easily exaggerated by one or two log instances for infrequent categories like \code{ai\_improve\_prompt} (0.17\%), \code{decompose\_task\_in\_prompt} (0.31\%), and \code{ai\_breakdown\_  intent} (0.65\%) (details in \cref{fig:code-distr-step} in Appendix \ref{sec:appendix-challenge}). 

\paragraphBold{Learning objectives of AI use skills across knowledge dimensions.}
\label{subsec:ai-lo}

To produce actionable insights for instructors, we examined both survey responses and log examples to identify AI use challenges that persist beyond classroom demonstrations and to distill concrete learning objectives.
From all 36 students' five post-survey answers to the questions ``What were the challenges during the tasks when you used GenAI'' and ``What do you think could help you further improve your performance with GenAI on these tasks?'', we derived learning objectives for \textbf{AI use skills} (\cref{tab:ai-use-los}).
We adopted the knowledge dimensions taxonomy to categorize AI use skills \cite{Krathwohl2002-nu, Wiggins2005-ev}. 
Using a thematic analysis procedure similar to \cref{subsubsec:qual-log-analysis}, we iteratively constructed the codebook via open coding \cite{strauss1998basics}, and we validated the codes with examples from log data (details in \cref{tab:ai-behaviors-lo-map}, Appendix \ref{sec:appendix-challenge}).

\paragraph{Differentiate Knowledge Dimensions}
In short, AI challenges that students encountered often stem from a lack of various AI use skills. 
In \cref{tab:ai-use-los}, we revealed necessary conceptual knowledge (what to know about AI), procedural skills (how to use AI), metacognitive skills (when to apply AI), and dispositional skills (keep engaging with AI).\footnote{There can be affective and motivational learning goals, such as self-efficacy and interest to explore. However, these goals are not manifested in our behavioral log data, and we discuss more about possible extensions in future work (\cref{subsec:limitation}).}
We provide some behavioral examples to help illustrate each AI use dimension. However, from our behavioral logs, we could not always identify which knowledge dimension exactly applies. For example, a student may know a particular way of using AI already, but they may not be able to retrieve it properly at the time they need to use it (available but not accessible \cite{Tulving1966-ai}). Especially when a student did not use AI, we may not conclude if it was because they did not notice the appropriate time to apply this knowledge (\textit{metacognitive}), did not have knowledge stored (\textit{conceptual}), failed to execute properly (\textit{procedural}), or lacked the persistence to keep trying (\textit{dispositional}). Future work may design more targeted assessments corresponding to each type of knowledge (\cref{subsec:limitation}).

\paragraph{Metacognitive.}
Students often struggled with deciding \textbf{\textit{when to involve the AI and what to delegate}}. 
As we analyzed HW1\&2 logs (after class instructions), we found multiple instances of students getting stuck in a loop of \code{explain\_error} and \code{generate\_code} without re-evaluating their approach (such as the novice's code edit sequence \cref{table:expert-novice-keyerror}). These failures stem from repeated \texttt{KeyError} exceptions --- as Gemini in Colab was not able to retrieve the correct column names from the dataframe, code generation without augmented context kept hallucinating nonexistent columns. %

\begin{table*}[]
\centering
\caption{
Comparison of experienced students vs. novice behaviors in resolving \code{KeyError: Column not found}. Experienced students identify missing context (e.g., actual column names) and fix issues using grounded prompts or direct code edits. Novices often rely on vague prompts or trial-and-error, leading to unresolved errors.
}
\vspace{-5pt}
\label{table:expert-novice-keyerror}
\small
\begin{tabular}{c | p{0.35\textwidth} | p{0.55\textwidth}}
\toprule
\textbf{Edit on} & \textbf{Technically Experienced Behavior} & \textbf{Novice Behavior} \\
\midrule
\textbf{Code} & 
\code{KeyError: total\_players} \newline
\hspace*{1em}Thinks aloud: "What's the actual column name?" \newline
\hspace*{1em}Checks \code{df.head()} output \newline
\hspace*{1em}Finds and copies \code{\#total\_players} \newline
\hspace*{1em}Updates code → error resolved
&
\code{KeyError: price} \newline
\hspace*{1em}Uses “Explain Error” → LLM suggests fixing to \code{Price} → applies fix → error resolved\newline %
\hspace*{1em}Repeats for other columns; some fixes succeed, others (e.g., \code{Interior Color}) fail \newline
\hspace*{1em}Continues blindly copy-pasting → unresolved errors accumulate \newline
\hspace*{1em}Eventually abandons data cleaning \\
\midrule
\textbf{Prompt} & 
\code{KeyError: Current Players} \newline
\hspace*{1em}Prompts: “Column is \code{current\_players}” \newline
\hspace*{1em}Error persists \newline
\hspace*{1em}Realizes true name: \code{\#current\_players} \newline
\hspace*{1em}Clarifies: “Use \code{\#current\_players} to match dataset” \newline
\hspace*{1em}Model regenerates → error resolved
&
\code{KeyError: gain} \newline
\hspace*{1em}Prompts: “Don't use columns that don’t exist” + pastes error \newline
\hspace*{1em}Thinks aloud: “I already said that, still doesn’t work...” \newline
\hspace*{1em}Tries variations of prompt → still fails and switches to different EDA question \newline
\code{KeyError: peak\_player} \newline
\hspace*{1em}Prompts again: “Don’t use non-existent columns” + pastes error \newline
\hspace*{1em}Thinks aloud: “I know it doesn’t exist… I don't know why” \newline
\hspace*{1em}Gives up and moves on \\
\bottomrule
\end{tabular}
\Description{Side-by-side examples comparing how experienced and novice students resolve KeyError issues. Experienced students inspect data and correct column names, while novices repeatedly use vague prompts and often fail to resolve the errors.}

\vspace{-10pt}
\end{table*}

This pattern could illustrate multiple missing AI use skills, such as a lack of metacognitive monitoring: the ability to recognize when a strategy is unproductive and to adjust. Students could avoid delegating AI to do the task that it has demonstrated unreliability and providing the correct column name themselves {(such as the experienced student's code-edit sequence \cref{table:expert-novice-keyerror})}, or alternatively ask AI to \code{breakdown} the debugging procedure and walk them through step-by-step, or add the exact column name as \code{context\_in\_prompt} {(such as the experienced student's prompt-edit sequence \cref{table:expert-novice-keyerror})}. Effective metacognitive use would involve pausing to decide, for example, whether to ask the AI for a structured plan (e.g., ``list all columns in my dataset'') before trial-and-error. %

\paragraph{Conceptual.}
Failures could stem from the lack of conceptual knowledge regarding the \textbf{\textit{AI's task space (what it can do) and context space (what information it can access)}}. 
For \code{task space}, most showed improvement in appropriate AI use (\cref{fig:ai_use_ratio_code_distr}); however, students may fail to learn with simple demonstrations. For instance, the instructor covered using AI to \code{improve\_prompt}, which appropriate use increased from 0\% to 100\% in post-demo (despite low frequency of 0.17\% per student). Nevertheless, some students still did not realize this use until the latter half of the semester: \citequote{HW3 post-survey}{now I found a smooth way to utilize GenAI -- ask it to also re-write my prompt before I use the prompt to generate the code.}.
If the demonstration did not cover some AI use, students may develop a wrong assumption. For example, the instructor did not cover \code{ai\_edit\_partial\_code} during the demo, and a student demonstrated struggles caused by the lack of this AI use knowledge in their post-survey: \quoteinline{you can't fine-tune parts of the result; instead, you have to regenerate the whole thing, and sometimes, parts that were correct before end up wrong.}

For \code{context space}, the instructor explicitly emphasized that Colab Gemini has \code{limited context space} during the demo, with a clear example of generation failure without providing explicit context to the model. However, in students' logs and post-surveys, we still observe that many students assumed the model could ``see'' their dataframe and column names without explicit context: \citequote{post-class survey}{I had to give small details of the code as well to the GenAI, I thought it already has taken the input that is there on the screen.}. 
{As shown in the novice's prompt-edit sequence in \cref{table:expert-novice-keyerror}), a novice might be able to tell that AI is hallucinating non-existent columns, but is still not able to realize they need to augment the context for AI.}
If the students recall about AI's limiting \code{context space}, they might recognize that the column name needs to be provided in the prompt to avoid hallucination.

\paragraph{Procedural.}
Just \textit{knowing} what AI needs is not enough, as students still need \textbf{\textit{skills to best interact with the AI}}. One student reflected on the challenge of providing appropriate contexts in prompts (procedural), even when knowing AI \code{context space} (conceptual): \citequote{post-class survey}{Most of the time, it was difficult to feed context because there were multiple sources to refer to}.
Similarly, one may know they should break down a complex goal without being able to accomplish it, and end up settling for downgraded goals: \quoteinline{I didn't really know how to write the prompt for the type of complex data analyses to generate the charts that I wanted, so I had to keep to simple analyses and charts.}
In the demo, the instructor exemplified good prompting strategies, such as writing \code{clear\_instruction}, \code{providing\_context}, and \code{decomposing\_task\_in\_prompt}.
However, post-class behaviors still leave 20\% of missing AI use as room for improvement (\cref{fig:ai_use_ratio_code_distr}).

\paragraph{Dispositional.}
Finally, students may lack \textbf{\textit{persistence and willingness to iterate with the AI}} and quit from frustration. Some learners disengaged quickly when outputs were erroneous or lengthy. 
In the afore-mentioned example of repeated \texttt{KeyError}, while AI failed to debug on its own and made repeated errors, some students' logged actions of \code{generate-code} and \code{explain-bug} were done within short time intervals of seconds, which indicates that the students did not persist in actually reading or understanding the error explanations or attempting to debug.

The lack of persistence may also result in early stops. For instance, when a student tried to add labels on top of a pie chart, the generated code produced errors, and simply pressing the ``explain error'' button did not fix the issue. The student stopped trying, accepting the basic pie chart without the label and commented \citequote{HW2 think-aloud}{seems like it either do the label or a pie}.
In comparison, success is shown when a student keeps trying to refine outputs until they achieve the intended goals, e.g., persisting through different AI mistakes to produce the ideal visualization: 

\begin{quote}
Given the above dataset, I need to see 5 years of pie charts comparing genre to release date. The goal is to understand which genres, as a percentage of all genres, were released each year \newline $\rightarrow$
[copied: ... df\_games.info() ] Regenerate the "genre to year" comparison again ...
All operations should be done within this dataset for all prompts going forward \newline $\rightarrow$
The graphic is too noisy to glean anything useful. Can we pick the top n games ... if a game has less than 3\% share ... we don't display it ...? \newline $\rightarrow$
You did the opposite of what I asked. ... display the genres OVER a certain threshold. 3\% was an example. Can you suggest a better way to find this statistically important threshold and then re output using it? \newline $\rightarrow$
Pie chart is not the answer. Can we try a histogram or bar chart? If those aren't wise, what would you recommend for comparing the genre to total players by year? \newline $\rightarrow$
regenerate to make the grouped bar chart \newline $\rightarrow$
Change this to output one such chart per year. There should be one chart for every year in the range.
\newline (HW2 log)
\end{quote}

\section{Discussion and Future Work}
\label{sec:discussion}

\subsection{Design and Pedagogy Implications}
\label{subsec:implications}
\label{subsec:challenges-in-the-wild}

\paragraphBold{Make AI mental models explicit.} 
Unlike simple chat interfaces {widely studied by existing works \cite{Chen2024-du, Sawalha2024-bz, Tassoti2024-tl, Nguyen2024-sd}}, AI integrated in a notebook exposes multiple affordances (code cells, outputs, error panes, menu buttons, etc.). 
We observed in the study that students transferred assumptions from more capable general-purpose AIs to their Colab working environment, miscalibrating what the embedded model could access or do. This produced confusion about AI \emph{task space} (what to ask) and \emph{context space} (what the model can “see”). 
In post-surveys, some students commented that they switched to other AI tools like ChatGPT because Gemini keeps producing bugs or have to prompt to exclude the exact column key every time.
As shown in \cref{table:expert-novice-keyerror}, a novice prompted the model ``Don't use nonexistent columns, use what you have in the datasheet'' after three repeated \texttt{KeyError} and commented ``Perplexity can do that,'' assuming that Colab Gemini would have access to the imported dataframe like other systems. It took them more than five repeated failures to realize that they should adjust their mental model for the system at hand, highlighting the need for explicit conceptual models and metacognitive abilities. 

{Similar issues exist in the long history of research on mental models of software systems \cite{Council1987-qs}.}
On the design side, systems with GenAI assistants should make their task and context scope more legible (e.g., what files and variables are visible to GenAI, what operations are supported, etc). Teaching “what to ask this specific AI” and “what this specific AI knows” is likely getting obsolete as models and systems rapidly evolve \cite{Ma2025-wf}. More importantly, instructions are needed for \textit{metacognitive abilities to adjust mental models to different AI system capabilities}, and methods such as self-reflection may help expand AI literacy to metacognitive skills \cite{Kumar2024-ck}.

\paragraphBold{Facilitate evaluation by providing less information and scaffolding verification.}
As we found in \cref{subsec:class-effect}, using AI to gain additional, evaluative information was a consistent challenge. Asking AI to explain code, bugs, or critique outputs remained limited, despite its potential benefit to improve understanding. We especially observe that students may not persist in reading convoluted output, reflecting reluctance or difficulty in engaging with evaluative or complex information. While there are one-click affordances (explain error buttons) that aim to make it easier to start evaluation, the cognitive load of seeing a lengthy error explanation often creates resistance or reduces the chance for students to learn critical evaluation skills. 

On one hand, we need instructions on \textit{dispositional skills to help humans stick to evaluation and mitigate the risk of overreliance}. On the other hand, in terms of feedback design for novice students or users, \textit{less is more.}
Explicit instructions that focus exclusively on evaluating LLM outputs may alleviate this cognitive load in a complex task environment \cite{Ma2024-gy}. 
Additionally, it is possible to lean on the rich log data to trigger real-time hints and output style changes. For example, repeated “fix it” after \texttt{KeyError} in short intervals suggests struggles of evaluation, so instead of providing a full code fix saying ``If your column name is X, the code will look like Y [lengthy code] [lengthy explanation],'' generating a diagnosis plan or hint for the user one step at a time might be more effective, e.g., ``Run \texttt{df.columns} and check the output. Copy-paste the exact column name you want to use in code.'' 

In terms of design, generating UIs \cite{Chen2025-cg} instead of lengthy text may help alleviate the evaluation burden. Designs that intentionally insert intermediate checkpoints can help decompose big evaluation plans into more manageable chunks, or frontload evaluation by asking users to provide requirements before generation, e.g., in the form of plan skeletons and verification checklists \cite{Kazemitabaar2024-sb}. Pedagogy can mirror this by requiring learners to specify success criteria and requirements before requesting code \cite{Ma2025-wf}.

\subsection{Limitations and Future Work}
\label{subsec:limitation}

\paragraphBold{Going deeper: ground in assessments and controlled experiment to study factors influencing effective LLM use}

Our analyses suggest that students' technical experience significantly predicts performance during LLM use (similar to prior literature \cite{Chen2024-du}), whereas LLM and communication experience factors did not. Additionally, we showed that the technical experience gap was reduced by LLM use under time pressure.
Other prior work also shows that LLMs can narrow gaps between novices and experts \cite{Pickering2025-uj}, or even impede expert performance under certain conditions \cite{Becker2025-ho}. 
{While we did not experimentally isolate the effect for time-on-task, we provide preliminary evidence that time constraints may influence the effectiveness of LLM use. The different results of whether LLM closes the experience gap or not in bi-weekly assignments versus tasks in lecture highlight that technical experience is not the only factor shaping effective LLM use (\cref{subsec:result-lmer}).}

Our sample included 36 students in authentic classroom contexts, which ensured ecological validity. However, it is a small sample that constrained our ability to derive conclusive subgroup analysis and isolate factors, such as undisclosed external AI use and overlapping experience dimensions. For example, Python and data science proficiency were moderately correlated in our sample, so we could not examine their independent effects on effective LLM use. 
Our log analyses revealed the importance of AI use strategies, which point toward the need to define and evaluate “AI use experience” more directly, as what we have attempted in \cref{subsec:ai-lo}.
We also measured affective factors (motivation, self-efficacy) in self-reported survey items (Appendix \ref{sec:appendix-surveys}), but none significantly predicted performance. 
Experience scores in our work were operationalized as self-reported survey items, which, despite being common practice \cite{Chen2024-du, Lum2025-sx, Chen2024-gx}, have some known issues such as miscalibration of self-perceived experience \cite{Kruger1999-io}. We partially validated our technical experience measure with the performance in HW0 tasks without using LLMs; however, future studies could recruit more distinct participant groups or employ more rigorous pretests. %
More robust performance-based assessment and controlled experiments are needed to disentangle the interplay between more nuanced factors such as domain experience, implementation experience, AI-use experience, affect, and situational factors like time pressure.

\paragraphBold{Going broader: extend to more diverse task contexts. }

The tasks employed in our study were appropriate for classroom learning but were not ``LLM-hard:'' current models could achieve high accuracy. Future work could incorporate more complex tasks that better reveal the value of AI use skills and enable observations on how learners adapt when models struggle. 

While the learning objectives for AI use skills (\cref{tab:ai-use-los}) were derived from programmatic data analysis tasks, they should extend beyond our study context. Many objectives are relevant to problem-solving and higher-order thinking skills, such as clarifying the task space of any tool, communicating requirements with decomposition, and verifying results. Future research could examine whether instruction on AI use transfers to more general competencies, and explore how to evaluate AI use experience more systematically (as self-rated LLM experience may not be appropriate).

Finally, we chose Google Colab for the class because it embeds a free, in-house Gemini, which enabled us to explore diverse behavioral AI use patterns in our research questions. 
Compared to prior work that analyzes linear chat logs~\cite{lee2022coauthor}, our setup in Google Colab involves multiple LLM-powered functions with different affordances and model access, resulting in more nuanced observations about when, how, and why students turn to AI support (e.g., comparing behaviors between AI-as-chat versus AI-as-inline-tool).
However, our data collection relied on a custom logging script in the browser, which can be less robust than existing Jupyter notebook logging toolkits like JELAI \cite{Valle-Torre2025-zi}. For example, our approach may capture an auto-completion as users' code edits (although infrequent), and Colab's frequently changing interface raises concerns about long-term replicability. Future work could build on established logging toolkits to improve data quality and generalizability.

\section{Conclusion}
\label{sec:conclusion}

LLMs promise to broaden access to technical work, yet our study shows that not everyone benefits equally. Technical experience remained the strongest predictor of success in homework, and more technically experienced students engaged with AI more strategically than novices. Behavioral log analysis highlighted that experienced students provide clearer context, decompose tasks, and strategically use AI, while novices tend to rely on AI reactively in the face of challenges. While demonstrations and extended time helped students adopt clearer prompting practices, evaluative skills remained challenging. 
These findings suggest that effective AI use needs to cultivate durable competencies in multiple dimensions. %
We translate these insights into design and pedagogy implications for AI literacy: explicate mental models, clarify before generating, and scaffold verification. 
Ultimately, effective AI use is not just better prompts --- it is about better problem solving.

\begin{acks}
Thanks to all the participants for the study, and thanks to all LearnLab members who provided feedback on this work. 
Thanks to the National Science Foundation (award CNS-2213791, 2414915) and Google Academic Research Award for partial support of this work.

\vspace{10pt}
\noindent \textbf{Acknowledgment of LLM Use.}
We acknowledge the use of ChatGPT to assist with language editing, summarization, and restructuring of text during manuscript preparation. 
Claude was used to assist the analysis of log data; implementation details are included in \cref{subsubsec:qual-log-analysis}, prompts and settings are attached in Appendix \ref{sec:appendix-prompt}.
Copilot was used to assist the visualization and code writing to process and clean our dataset, and Gemini in Google Colab was used to prepare the homework materials and reference solutions. 
\end{acks}

\bibliographystyle{ACM-Reference-Format}
\bibliography{paperpile,ref}

\clearpage\newpage

\appendix

\onecolumn

\section{Codebook for task challenge and AI use success and failure behaviors}
\label{sec:appendix-challenge}

\begin{table}[htbp]
\centering
\caption{Processes and task challenges in workflows. Effective task completion is a combination of both correctness and efficiency.}
\label{tab:task-challenges}
\begin{tabular}{p{1cm}p{4cm}p{3cm}p{6cm}}
\toprule
\textbf{Process} & \textbf{Process Description} & \textbf{Task Challenge} & \textbf{Task Challenge Description} \\
\midrule
{Form\newline Input} 
& \multirow{2}{=}{\parbox{4cm}{Expressing intent through correct Python code \& producing desirable output for data science tasks}} 
& Form Correct Intent for Task & Effectively knowing what to do (e.g., edit code, refine visual, clarify concept, break down complex goal, etc.). \\
& & Express Intent in Code & Effectively generate code (for the data science task) for the right intent to produce desirable outputs. \\
\midrule
{Process Output} 
& \multirow{2}{=}{\parbox{4cm}{Making sense of code output; checking correctness and alignment with goals}} 
& Understand Output (Execution / Text) & Effectively read and understand code execution output, such as error messages, displayed images, tables, and text. \\
& & Assess Output & Effectively evaluating whether the output is correct and aligned with task goals. \\
\bottomrule
\addlinespace
\end{tabular}
\Description{Table summarizing student-reported challenges when using GenAI, such as inaccurate outputs, unclear explanations, and difficulty refining prompts.}
\end{table}

\begin{table*}[htbp]
\centering
\vspace{-20pt}
\caption{Mapping between the success and failure AI use behaviors (code frequency in \Cref{fig:code-distr-step}) and corresponding missing AI use skills (learning objectives).}
\small
\label{tab:ai-behaviors-lo-map}
\begin{tabular}{p{0.2\linewidth}p{0.27\linewidth}p{0.27\linewidth}p{0.18\linewidth}}
\toprule
\textbf{Success AI use behaviors} & \textbf{Failure AI use behaviors} & \multicolumn{2}{l}{\textbf{AI use challenges due to missing AI use skills}}  \\

\midrule
\multicolumn{3}{l}{Task Challenge: Form Correct Intent for Task} & \\
\midrule
\texttt{ai\_explain\_concepts}, \texttt{ai\_suggest\_steps\_or\_plan}, \texttt{ai\_breakdown\_intent}. & 
Stuck as they ``don’t know what can I ask AI.'' \newline
Did not use AI to assist in forming intent. & 
\textbf{[Conc-Task]}: know AI can explain concepts, suggest/plan steps, and decompose tasks. \newline
\textbf{[Proc]}: when prompting AI. & 
\textbf{[Meta]}: monitor when to use AI. \newline 
\textbf{[Disp]}: actively explore and persist through ambiguity \\ %
\addlinespace

\midrule
\multicolumn{3}{l}{Task Challenge: Express Intent in Input (Code)} & \\
\midrule
\texttt{clear\_instruction}, \newline 
\texttt{provide\_context}, \newline 
\texttt{decompose\_task\_in\_prompt}. \newline
\texttt{ai\_improve\_prompt}, \texttt{ai\_generate\_code}, \texttt{ai\_edit\_partial\_code}. & 
Fail to clearly specify intent. \newline
Fail to include necessary context $\rightarrow$ hallucinations. \newline
Fail to decompose overly complex intent. \newline
Did not use AI to help formulate the prompt. \newline
Fail to instruct AI to edit code. & 
\textbf{[Proc-Clear/Context/Decompose]}: use a prompt that clearly specifies intent, contains context, and has a decomposed goal. \newline 
\textbf{[Conc-Context]}: know AI needs context augmentation. \newline 
\textbf{[Conc-Task]}: know AI can improve prompts and edit partial code. & 
\textbf{[Meta]}: monitor when to use AI. \newline 
\textbf{[Disp]}: patience to re-prompt with more context, instead of repeating the sub-optimal prompt. \\
\addlinespace

\midrule
\multicolumn{3}{l}{Task Challenge: Understand Output (Execution / Text)} & \\
\midrule
\texttt{ai\_explain\_code}, \texttt{ai\_explain\_bug}, \newline
Read and {interpret AI outputs}.
& 
Did not use AI to help explain code. \newline
Did not try to understand the text and just passively copy-pasted. & 
\textbf{[Conc-Task]}: know AI can explain APIs, code, and bugs. \newline 
\textbf{[Proc]}: when prompting AI. & 
\textbf{[Meta]}: when to use AI. \newline 
\textbf{[Disp]}: sustained effort to interpret long outputs. \\
\addlinespace

\midrule
\multicolumn{3}{l}{Task Challenge: Assess Output} & \\
\midrule
\texttt{ai\_critique\_output}. & 
Did not use AI to evaluate output. \newline
Accepted AI critique without verification. & 
\textbf{[Conc-Task]}: know AI can critique output. \newline 
\textbf{[Proc]}: when prompting AI. & 
\textbf{[Meta]}: when to use AI. \newline
\textbf{[Disp]}: patience to carefully use AI to evaluate. \\
\bottomrule
\end{tabular}
\Description{Examples linking log excerpts to specific AI-use learning objectives, illustrating conceptual, procedural, metacognitive, and dispositional challenges.}
\end{table*}

Norman's Stages of Action~\cite{tenner2015design} describes the user cognitive process as a sequence of \emph{forming the goal} and \emph{the intention}, \emph{specifying} and \emph{executing} an action, \emph{perceiving} and \emph{interpreting the world state}, and \emph{evaluating the outcome}.
Inspired by this theoretical framework, we thematically coded the survey answers for ``What were the challenges for you during the tasks when you did \emph{not} use GenAI?'' from the HW0 post-survey, focusing on task challenges that exist even without AI use and could be addressed with effective AI use. %

As shown in \cref{tab:task-challenges}, students reported challenges throughout \code{forming intent}, \code{writing code}, \code{understanding output}, and \code{assessing output}.
For example, students reported that they \citequote{\code{form intent}}{did not know where to start at all}, had challenges in \citequote{\code{write code}}{putting the right code and iteratively coding on it}, \citequote{\code{understand output}}{interpreting the errors}, and \citequote{\code{assess output}}{knowing what kind of answer is good}.
Students who did not have any programming or data science background reported that they \quoteinline{do not know what to do} and \quoteinline{have no idea if I'm doing it right or even understand what I'm doing.} Even students with some programming experiences described the process as \quoteinline{trying to build a house with just a hammer and some nails} and \quoteinline{working blindfolded in a room full of puzzle pieces}. %

\section{Study Method Details}
\label{appen:method-detail}

\begin{figure}[!ht]
\centering
\begin{minipage}[t]{0.5\textwidth}
\vspace{-6cm}
Here we list the data source details for each analysis method (\cref{sec:method}):
\begin{itemize}
    \item \textbf{RQ1} (\cref{sec:skill-grade}): 
    \begin{itemize}
        \item Mann-Whitney U test (36 students, HW0, 36 submissions), 
        \item LMER (36 students, HW1-4, 144 submissions),
    \end{itemize}
    \item \textbf{RQ2} (\cref{sec:ai-behavior}): 
    \begin{itemize}
        \item AI use behaviors codebook (\cref{tab:ai-codebook}) construction (10 students, 5 screen videos + logs from HW0, 5 think-aloud videos + logs from HW1-4),
        \item Annotated log analysis (28 students, HW0-2, 36 submissions),
        \item See method details and justifications in \cref{subsec:log-analysis},
    \end{itemize}
    \item \textbf{RQ3} (\cref{sec:skill-to-teach}): 
    \begin{itemize}
        \item AI use skills codebook (\cref{tab:ai-use-los}) construction (36 students, HW0-4, 180 post-surveys), 
        \item Annotated log analysis (28 students, HW0-2, 36 submissions),
    \end{itemize}
    \item Qualitative examples could be selected from all students logs / videos / surveys from all homework.
\end{itemize}
\end{minipage}
\hfill
\begin{minipage}[t]{0.35\textwidth}
  \centering
  \includegraphics[trim=2cm 17.5cm 11.5cm 2.1cm, clip, width=\linewidth]{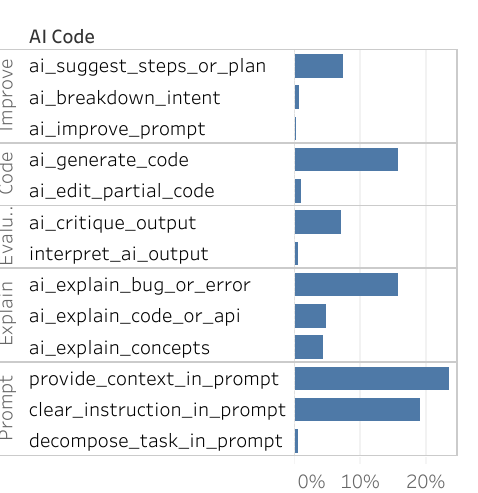}
  \caption{Distribution of AI use code frequency averaged per student, aggregated across all steps and clustered by AI usage code types. Refer to \cref{tab:ai-codebook} for definitions.}
  \label{fig:code-distr-step}
\Description{Bar charts showing the overall frequency of each AI usage behavior across logs. Providing context and requesting steps are frequent, while prompt refinement and partial code edits are rare.}
\end{minipage}

\end{figure}

\section{Survey Instruments}
\label{sec:appendix-surveys}

Here, we summarize survey instruments administered during the study. Questions are grouped into (i) shared items common across multiple surveys, and (ii) unique items included only in the Pre Survey, HW0, HW1–HW4, or Final Post Survey. Some survey items are drawn from existing measures such as NASA Task Load Index (TLX) for cognitive load \cite{Hart1988-kv}.

\subsection{Pre Survey Only Questions}
\begin{itemize}
    \item Demographics: ``How old are you?''; ``What is your gender?'' (Male/Female/Non-binary/Prefer not to say).
    \item Background: ``What major did you study for undergraduate/college?''; ``What program are you studying now at CMU?''.
    \item Expertise: ``How would you assess your level of expertise in the following areas?'' (Programming, Communication, GenAI/LLM, Data cleaning, EDA, ML, Storytelling, Car industry [dataset domain], Video game [dataset domain]; 5-point scale).
    \item Interest: ``How would you assess your level of interest in the following areas?'' (same domains, 5-point scale).
    \item Frequency: ``How frequently do you use GenAI/LLM for Programming, Communication, Data cleaning, EDA, ML, Storytelling?'' (Never–Daily).
    \item Experience: ``How many years of experience do you have with Excel, R, Python, Jupyter/Colab, pandas, Altair, scikit-learn?'' (0, 1–2, 3–5, 6–10, >10).
    \item Confidence: ``How confident are you in your ability to do the following tasks on your own?'' (Programming, Data cleaning, EDA, ML, Storytelling).
    \item GenAI ability: ``How confident are you in GenAI/LLM's ability to do the following tasks?'' (same domains, 5-point scale).
    \item Attitudes: ``I am confident in my ability to communicate to GenAI/LLM so that it does what I want.'' (5-point Likert).
    \item Expectations: ``What do you want to learn from this class?''.
\end{itemize}

\subsection{Shared Questions Across Post Surveys}

\begin{itemize}
    \item \textbf{GenAI tool usage.} ``What GenAI did you use for this homework? (Colab Gemini, ChatGPT, Copilot, Perplexity, Claude)''.
    \item \textbf{Perceived difficulty.} ``How mentally demanding/difficult was the task?'' (5-point Likert from ``Not difficult at all'' to ``Extremely difficult'').
    \item \textbf{Effort.} ``How hard did you have to work for the task?'' (5-point Likert from ``I do not need to work hard at all'' to ``I need to work extremely hard'').
    \item \textbf{Frustration.} ``How irritated, stressed, and annoyed did you feel during the task?'' (5-point Likert).
    \item \textbf{Interest.} ``How interesting/engaging was the task?'' (5-point Likert).
    \item \textbf{Self-evaluation.} ``How successful were you in performing the task? How satisfied were you with your performance?'' (5-point Likert from ``Not successful/satisfied at all'' to ``Extremely successful/satisfied'').
    \item \textbf{Confidence.} ``How confident are you that you did well on the task?'' (5-point Likert).
    \item \textbf{Peer help.} ``How much did you ask your peer/classmate for help with the task?'' (5-point Likert from ``Not at all'' to ``All the time'').
    \item \textbf{Open-ended reflections.} 
    \begin{itemize}
        \item ``What was challenging for you during the tasks when you used GenAI?''.
        \item ``How did you use GenAI during the tasks?''.
        \item ``Did you find using GenAI helpful during the tasks? Why or why not?''.
        \item ``What do you think could help you further improve your performance with GenAI on these tasks?''.
        \item ``What help did you offer to or receive from peers/classmates? What questions do you ask them instead of Google or GenAI?''.
    \end{itemize}
\end{itemize}

\subsection{HW0 Post Survey Only Questions}
\begin{itemize}
    \item Task assignment: ``Which tasks did you do with GenAI?''; ``Which tasks did you do without GenAI?''.
    \item Link: ``Link to Zoom recording''.
\end{itemize}

\subsection{HW1–HW4 Post Survey Only Questions}
\begin{itemize}
    \item Each survey contextualized task focus:
    \begin{itemize}
        \item HW1: ``Part 1. Data cleaning''.
        \item HW2: ``Part 2. Exploratory data analysis''.
        \item HW3: ``Part 3. Machine learning''.
        \item HW4: ``Part 4. Data-driven storytelling''.
    \end{itemize}
\end{itemize}

\subsection{Final Post Survey Only Questions}
\begin{itemize}
    \item Retrospective self-assessment of expertise, interest, and confidence in self vs.\ GenAI ability across domains: E.g., ``How would you assess your level of expertise in the following areas?'' (same as Pre Survey, 5-point scale).
    \item Trust, communication, and AI self-efficacy statements: E.g., `Generally, I trust GenAI/LLM'' (same as Pre Survey). %
    \item Open-ended retrospective variants:
    \begin{itemize}
        \item ``What was challenging for you during the tasks when you used GenAI? Any similarities or differences across the assignments?''.
        \item ``How did you use GenAI during the tasks? Any similarities or differences?''.
        \item ``What did you learn from this class? What do you wish that you could learn more about?''.
    \end{itemize}
\end{itemize}

\section{Composite Experience Measures}
\label{sec:appendix-competency}

\subsection{Grouping Construction Process}
\begin{figure}[t]
  \centering

  \begin{subfigure}[t]{0.52\linewidth}
    \centering
    \includegraphics[width=\linewidth]{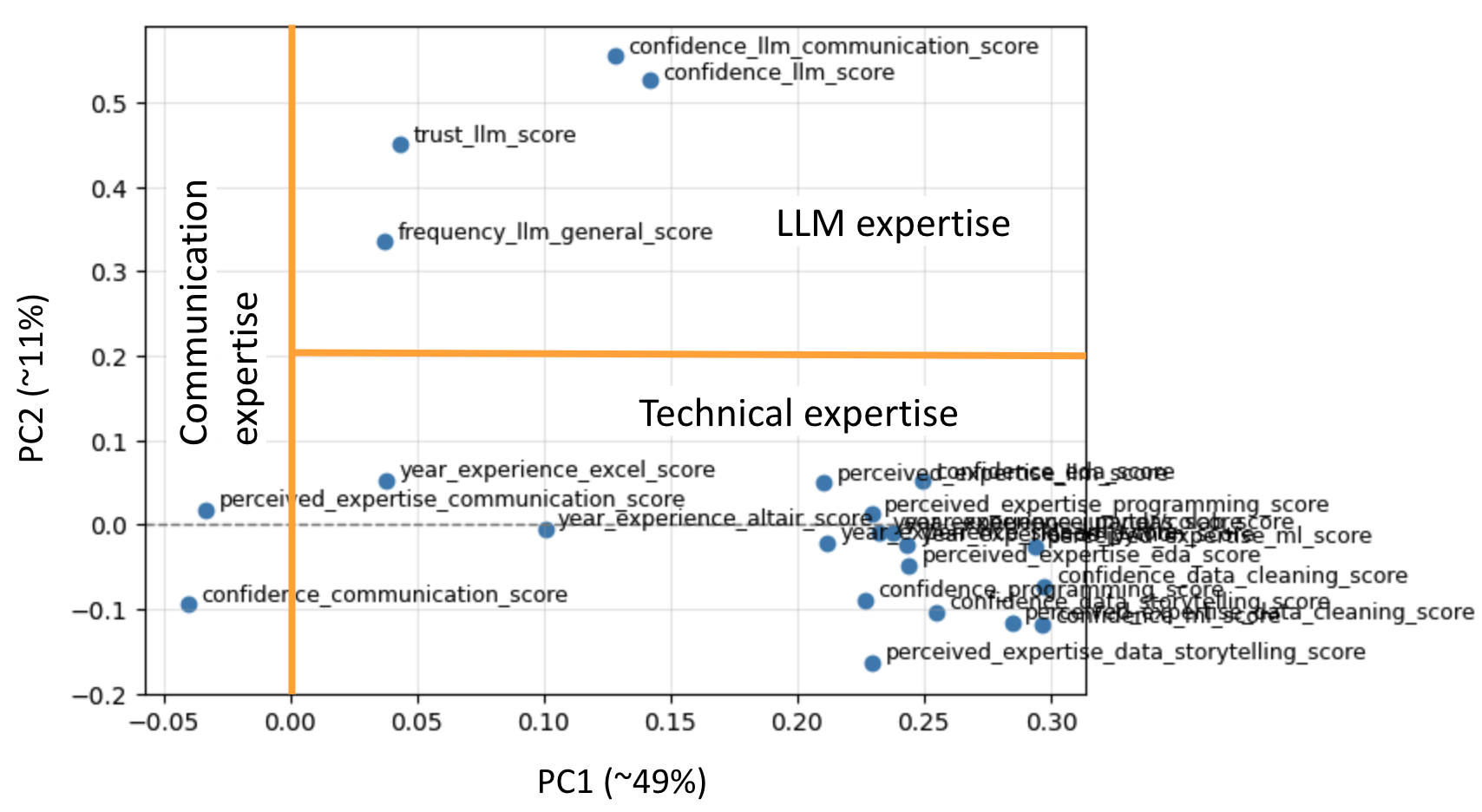}
    \caption{PCA of self-reported experience measures.}
    \label{fig:pca}
    \Description{PCA scatterplot of self-reported experience measures, with each variable positioned by its loading on PC1 and PC2. Technical experience variables cluster along the positive PC1 axis, LLM-related variables along positive PC2, and communication measures in the negative regions of both axes.}
  \end{subfigure}
  \hfill
  \begin{subfigure}[t]{0.45\linewidth}
  \vspace{-5cm}
    \centering
    \caption{Pairwise-correlation matrix and VIF scores for composite experience measures.}
    \label{tab:pairwise-corr-vif}
    \begin{tabular}{r|ccc|c}
      \toprule
       & Tech & LLM & Commun. & VIF \\
      \midrule
      Tech    & 1.00  & 0.41  & -0.23 & 1.22 \\
      LLM     & 0.41  & 1.00  & -0.28 & 1.26 \\
      Commun. & -0.23 & -0.28 & 1.00  & 1.10 \\
      \bottomrule
    \end{tabular}
    \Description{Variance Inflation Factor analysis and pairwise correlation matrix showing low multicollinearity among technical, LLM, and communication experience variables.}
  \end{subfigure}

  \caption{Overview of self-reported experience measures: (a) PCA and (b) pairwise correlations and VIF scores.}
  \label{fig:experience-overview}
\end{figure}

The experience variables are rationally constructed to enable interpretability and are also empirically justified with correlation analysis and principal component analysis (PCA) \cite{Greenacre2022-rz}. As shown in Figure~\ref{fig:experience-overview}a, PC1 captures the most variance ($\sim$49\%), aligning with technical experience; PC2 explains additional variance related to LLM experiences ($\sim$11\%); Communication-related measures cluster around the negative region of both PC1 and PC2. Note that given the low sample size and limited statistical power, we used PCA only to guide and visually verify the clustering of variables. 
We further validated the independence of these experience variables using correlation analysis and varience inflation factors (VIF) \cite{Marcoulides2019-wz}, as shown in Figure~\ref{fig:experience-overview}b. The lack of multicollinearity enabled further regression analysis of the experience variables in \cref{subsec:result-lmer}.

\subsection{Grouping Items}
\subsubsection*{Technical experience (composite of programming, data science, and tools)}
\begin{itemize}
  \item How would you assess your level of expertise in the following areas: [Programming; Data cleaning; Exploratory data analysis; Machine learning; Data-driven storytelling] \newline
  \emph{Scale:} Novice (1); Beginner (2); Intermediate (3); Advanced (4); Expert (5)
  \item How many years of experience do you have in using the following tools? [Python; Excel; Jupyter/Colab notebook; pandas; altair; scikit-learn] \newline
  \emph{Scale:} 0 (1); 1--2 years (2); 3--5 years (3); 6--10 years (4); >10 years (5)
  \item How confident are you in your ability to do the following tasks on your own? [Programming; Data cleaning; Exploratory data analysis; Machine learning; Data-driven storytelling] \newline
  \emph{Scale:} Not confident at all (1) -- Extremely confident (5)
\end{itemize}

\subsubsection*{LLM experience}
\begin{itemize}
  \item How would you assess your level of expertise in the following areas: [GenAI/LLM] \newline
  \emph{Scale:} Novice (1); Beginner (2); Intermediate (3); Advanced (4); Expert (5)
  \item How frequently do you use GenAI/LLM for: [In general] \newline
  \emph{Scale:} Never (1); Rarely (2); Monthly (3); Weekly (4); Daily (5)
  \item Please rate the following in terms of how much you agree or disagree with each statement: [I am confident in my ability to use GenAI/LLMs; I am able to have LLMs complete the tasks I instruct; I am confident in my ability to communicate to LLM so that it does what I want; Generally, I trust GenAI/LLM] \newline
  \emph{Scale:} Strongly disagree (1) -- Strongly agree (5)
\end{itemize}

\subsubsection*{Communication experience}
\begin{itemize}
  \item How would you assess your level of expertise in the following areas: [Communication] \newline
  \emph{Scale:} Novice (1); Beginner (2); Intermediate (3); Advanced (4); Expert (5)
  \item Please rate the following in terms of how much you agree or disagree with each statement: [I am confident in my communication ability in general] \newline
  \emph{Scale:} Not confident at all (1) -- Extremely confident (5)
\end{itemize}

\section{Prompt for Log Analysis}
\label{sec:appendix-prompt}

A \texttt{temperature} of 0 is used with the \texttt{model} \texttt{claude-sonnet-4-20250514} and \texttt{max\_tokens} of 20000, in the following two steps as a Chain-of-Thought:

\subsection{Segmentation Prompt}

\begin{framed}

\noindent\textbf{Task}\\
Turn a chronological JSON log of events from Google Colab (students completing a data-science task with help from the Gemini model) into episode segmentation.

\vspace{0.5em}
\noindent\textbf{Key Definitions:}\\
\textbf{Episode}: A contiguous sequence of actions (form intent, form input, understand output, assess output) driven by a single \textbf{overall intent} (e.g., ``remove duplicates in data column X'', ``fit a regression model to predict Y,'' ``creating visualization to analyze the trend of Z''). An episode ends when the student succeeds, gives up, or pivots intent. 

\vspace{0.25em}
\noindent\textbf{Decision boundary for different intent (and split into new episode):}
\begin{itemize}
  \item a different EDA question is a different intent
  \begin{itemize}
    \item a different plot is a different intent, even if it visualizes the same relationship or answers the same EDA question
  \end{itemize}
  \item cleaning a different column is a different intent
  \begin{itemize}
    \item removing outliers is different intent from making a column dtype numeric, even if they are both cleaning the same column
  \end{itemize}
\end{itemize}

\vspace{0.25em}
\noindent\textbf{Episode Steps:} each episode includes the following four steps; early stops/failures may not reach beyond step 1:
\begin{enumerate}
  \item \textbf{Form Intent}: Deciding what needs to be done to progress the task (e.g., remove outliers in a column, create a bar chart, refine a visual by sorting by count, clarify concept, break down a complex goal). Each unique intent should start a new episode. Writing comments and prompting model are common behaviors that can be used to infer intent.
  \item \textbf{Form Input}: Translating intent into input (Python code in \texttt{edit\_code} or an AI prompt in \texttt{converse} or \texttt{generateCode}) that can generate the desired result.
  \item \textbf{Understand Output}: Reading and interpreting the system's response (\texttt{llm} output in \texttt{event\_response}, \texttt{executed\_output}, error messages in \texttt{executed\_error}, plots, tables, or text).
  \item \textbf{Assess Output}: Evaluating whether the output is correct, useful, and aligned with the intended goal.
\end{enumerate}

\vspace{0.5em}
\noindent\textbf{Instructions:}\\
\textbf{1. Segment logs into episodes:}
\begin{itemize}
  \item Start a new episode when you observe a clear change in intent (see decision boundary above), \textbf{OR} if there is a time gap $\geq$ 15 minutes between consecutive events (use \texttt{datetime\_timestamp}).
  \item Ensure all log events are covered. Episodes should be consecutive, from the first log to the last.
  \item Prefer fewer coherent episodes unless there is a clear goal shift or long gap.
  \item Each episode must have \texttt{timestamp\_start} and \texttt{timestamp\_end} taken from the episode's first/last event.
\end{itemize}

\noindent\textbf{2. Identify the episode's overall intent:}
\begin{itemize}
  \item Infer from code, prompts, and outputs (e.g., remove outliers in a column, create a bar chart, refine a visual by sorting by count).
  \item Write 1--2 sentences describing what the user was trying to do and what happened, inferred from the log events.
\end{itemize}

\vspace{0.5em}
\noindent\textbf{Output Format (JSON array of episode objects):}
\begin{verbatim}
{
  "episode_id": 1,
  "timestamp_start": "datetime_timestamp",
  "timestamp_end": "datetime_timestamp",
  "events": ["evt_0", "evt_1", ... all event_ids in the episode],
  "overall_intent": "inferred user intent from logs"
}
\end{verbatim}

\end{framed}

\subsection{Step-wise Annotation Prompt}

\begin{framed}

\noindent\textbf{Task}\\
Turn an episode of chronological JSON log events from Google Colab (students completing a data-science task with help from the Gemini model) into episode steps with fine-grained actions and returns. Then annotate AI usage and its success/failure in the episode.

\vspace{0.5em}
\noindent\textbf{Input}\\
Each event contains fields like: \texttt{event\_id}, \texttt{event\_name}, \texttt{event\_category}, \texttt{event\_content}, \texttt{event\_response}, and \texttt{datetime\_timestamp}. AI use behaviors are recorded in events with \texttt{event\_type} = \texttt{converse} or \texttt{generateCode}.

\vspace{0.5em}
\noindent\textbf{Key Definitions}\\
\textbf{There are four different types of steps in an episode:}
\begin{enumerate}
  \item \textbf{Intent} -- forming a correct goal
  \item \textbf{Input} -- expressing the goal clearly in code or prompt to generate desired output
  \item \textbf{Understand} -- interpreting output/errors
  \item \textbf{Assess} -- judging output correctness \& alignment with goal
\end{enumerate}

\noindent\textbf{Users may display different AI-use behaviors in different steps of the episode:}
\begin{itemize}
  \item \textbf{intent-step}: asks AI to explain concepts or suggest steps; asks AI to break down steps.
    \begin{itemize}
      \item \texttt{ai\_explain\_concepts}
      \item \texttt{ai\_suggest\_steps\_or\_plan}
      \item \texttt{ai\_breakdown\_intent}
    \end{itemize}
  \item \textbf{input-step}: asks AI to improve/clarify the prompt; ask AI to edit code or generate code
    \begin{itemize}
      \item \texttt{ai\_improve\_prompt}
      \item \texttt{ai\_edit\_partial\_code}
      \item \texttt{ai\_generate\_code}
    \end{itemize}
  \item \textbf{understand-step}: asks AI to explain code or error, \texttt{press\_explain\_error}, \texttt{press\_explain\_code}, or clarifies APIs 
    \begin{itemize}
      \item \texttt{ai\_explain\_bug\_or\_error}
      \item \texttt{ai\_explain\_code\_or\_api}
    \end{itemize}
  \item \textbf{assess-step}: asks AI to critique output; queries alignment with goals/insight quality
    \begin{itemize}
      \item \texttt{ai\_critique\_output}
    \end{itemize}
  \item \textbf{step-agnostic-ai-use}: good AI use can happen across steps (whenever the user is using AI)
    \begin{itemize}
      \item \texttt{clear\_instruction\_in\_prompt}
      \item \texttt{provide\_context\_in\_prompt}
      \item \texttt{decompose\_task\_in\_prompt}
    \end{itemize}
\end{itemize}

\vspace{0.25em}
\noindent\textbf{Behavior Examples}\\
The user may encounter challenges at any step in the episode.

\noindent\textbf{Some example behaviors that demonstrated challenges in each step:}
\begin{itemize}
  \item \textbf{intent-step}: user stuck (``don't know what to do''), failure to form intent (e.g., ``what else can I do?'' / ``give me steps''), wrong intent (e.g., asks for adding ``\$'' instead of making a price column numeric), cannot decompose a complex goal.
  \item \textbf{input-step}: vague pronouns (``fix it'') without shared context, missing/incorrect context leading to hallucinations (e.g., \texttt{KeyError} of made-up columns), faulty code (syntax error, wrong API usage), \texttt{edit\_code} followed by \texttt{executed\_error}.
  \item \textbf{understand-step}: copy-paste generated code blindly, immediate ``fix it'' after \texttt{press\_explain\_error} without reading, repeating error loops with no adjustment.
  \item \textbf{assess-step}: accepts generation without verification (very short intervals), keeps ineffective chart/insight misaligned with goals.
\end{itemize}

\noindent\textbf{Some success behaviors for each step:}
\begin{itemize}
  \item \textbf{intent-step}: final intent is clear and correct (e.g., valid cleaning procedure, valid EDA question)
  \item \textbf{input-step}: input code/prompt are clear and correct (errors resolved)
  \item \textbf{understand-step}: demonstrated understanding via a concrete next action or time spent reading
  \item \textbf{assess-step}: result (data/chart/stats) is correct; or student requests targeted improvement based on assessment
\end{itemize}

\vspace{0.5em}
\noindent\textbf{Instructions}\\
\textbf{1. For each step of the episode, trace key events and summarize into Actions \& Returns with evidence:}
\begin{itemize}
  \item \textbf{Action}: initiate an attempt (e.g., edit code, enter prompt, send Gemini prompt, UI button like execute after editing code, \texttt{press\_explain\_error}, etc.)
  \item \textbf{Return}: immediate observable system feedback (stdout/stderr, exceptions, plots, metrics, AI response)
  \item Steps and evidence \texttt{event\_id}s must be in chronological order.
\end{itemize}

\noindent\textbf{2. For each step, annotate if a challenge was encountered, and whether AI was used strategically to overcome it:}
\begin{itemize}
  \item Use the challenge/success behavior examples to mark \texttt{encountered\_challenge}.
  \item If AI was used well, tag behaviors (e.g., \texttt{ai\_explain\_concepts}, \texttt{clear\_instruction\_in\_prompt}) and briefly explain how it helped.
  \item If not overcome, note which AI-use behaviors could have helped and why.
\end{itemize}

\noindent\textbf{3. Decide if the step was successful, and explain:}
\begin{itemize}
  \item \texttt{step\_success} can be true/false (partial progress or step not reached counts as false).
  \item Provide 2--3 sentences citing how the outcome is inferred from logs; include representative \texttt{event\_name}s and short excerpts from \texttt{event\_content}/errors/prompts or relevant timestamp gaps.
  \item Steps can be independent (e.g., wrong intent but correct understanding/assessment).
\end{itemize}

\vspace{0.25em}
\noindent\textbf{Visual-output assessment}\\
If the output involves a chart, evaluate the output code to judge whether the visualization answers the EDA question:
\begin{enumerate}
  \item Identify if there is visualization code (e.g., \texttt{alt.Chart}) intended to answer an EDA question. Evaluate the code even if the actual image is not available.
  \item Decide if the visualization is appropriate. If not, did the user refine it appropriately? Examples of assessment failures include:
    \begin{itemize}
      \item accepting an overly-busy chart (e.g., all genres when only top genres matter);
      \item using the wrong plot type (e.g., line instead of scatter for correlation);
      \item using a boxplot to show average rating (producing short lines rather than showing a range).
    \end{itemize}
\end{enumerate}

\vspace{0.5em}
\noindent\textbf{Output format}\\
Return output strictly as a JSON array of objects with these fields only:
\begin{verbatim}
{
  "step_id": 1,
  "step": "intent | input | understand | assess",
  "actions": [
    {"action": "event_name + short content excerpt", "summary": "what happened", 
     "evidence": "evt_0" | ["evt_0","evt_1"]}
  ],
  "returns": [
    {"return": "stdout|stderr|exception|plot|ai_response", "excerpt": "string", 
     "evidence": "evt_1" | ["evt_0","evt_1"]}
  ],
  "encountered_challenge": true | false,
  "ai_used_to_overcome_challenge": [] | ["ai_explain_concepts","clear_instruction_in_prompt"],
  "how_ai_helped": "" | "briefly explain ai_used_to_overcome_challenge",
  "ai_usage_that_couldve_helped": [] | ["provide_context_in_prompt"],
  "why_ai_could_help": "" | "briefly explain ai_usage_that_couldve_helped",
  "step_success": true | false,
  "reasoning": "how is the step & outcome inferred from logs"
}
\end{verbatim}

\end{framed}

\end{document}